\documentclass[floatfix,aps,pra,superscriptaddress,reprint]{revtex4-1}

\usepackage[T1]{fontenc}
\usepackage[utf8]{inputenc}
\usepackage{lmodern}
\usepackage{siunitx}
\usepackage{ulem}
\usepackage{enumerate}
\usepackage{amsmath}
\usepackage{amssymb}
\usepackage{siunitx}
\usepackage{physics}
\usepackage{bm}
\usepackage{graphicx}
\usepackage{listings}
\usepackage[usenames, dvipsnames]{color}

\newcommand{\Loic}[1]{{\color{black}#1}}

\newcommand{\ilan}[1]{{\color{black}#1}}

\graphicspath{ {images/}, {figures/} }

\begin{document}

\title{Tunable bandwidth and nonlinearities in an atom-photon interface\\ with subradiant states}

\author{Ilan Shlesinger}
\affiliation{Laboratoire Charles Fabry, Institut d'Optique, CNRS, Université Paris-Saclay, 91127 Palaiseau CEDEX, France.}

\author{Pascale Senellart}
\affiliation{Centre de Nanosciences et de Nanotechnologies, CNRS, Univ. Paris-Sud, Université Paris-Saclay, C2N - Marcoussis, 91460 Marcoussis, France.}

\author{Loïc Lanco}
\affiliation{Centre de Nanosciences et de Nanotechnologies, CNRS, Univ. Paris-Sud, Université Paris-Saclay, C2N - Marcoussis, 91460 Marcoussis, France.}
\affiliation{Université Paris Diderot, Paris 7, 75205 Paris CEDEX 13, France.}

\author{Jean-Jacques Greffet}
\email[E-mail: ]{jean-jacques.greffet@institutoptique.fr}
\affiliation{Laboratoire Charles Fabry, Institut d'Optique, CNRS, Université Paris-Saclay, 91127 Palaiseau CEDEX, France.}

\date{\today}

\begin{abstract}
Optical non-linearities at the single photon level are key features to build efficient photon-photon gates and to implement quantum networks. Such optical non-linearities can be obtained using an ideal two-level system such as a single atom coupled to an optical cavity. While efficient, such atom-photon interface however presents a fixed bandwidth, determined by the spontaneous emission time and thus the spectral width of the cavity-enhanced two-level transition, preventing an efficient transmission to bandwidth-mismatched atomic systems in a single quantum network. In the present work, we propose a tunable atom-photon interface making use of the direct dipole-dipole coupling of two slightly different atomic systems. We show that, when weakly coupled to a cavity mode and directly coupled through dipole-dipole interaction, the subradiant mode of two slightly-detuned atomic systems is optically addressable and presents a widely tunable bandwidth and single-photon nonlinearity.
\end{abstract}

\maketitle

\section{Introduction}

Quantum light  is a natural support of the quantum information \cite{knill2001}  for entanglement distribution in  quantum networks and long distance quantum communications~\cite{briegel1998,kimble2008}. In this context, efficient light-matter interfaces are key building blocks to store single photons in quantum memories, to manipulate them in order to build efficient photon-photon gates~\cite{shomroni2014,reiserer2014,hacker2016,volz2012} for local quantum computing or to implement quantum relays~\cite{bonato2010}. The manipulation or storage of single photons has been demonstrated both with atomic ensembles~\cite{ julsgaard2004, chaneliere2005, sayrin2015} or with individual atoms \cite{ specht2011, choi2008, korber2018}. The latter approach relies on the anharmonicity of a two-level system in a single natural atom or a solid-state artificial atom, and results in an optical non-linearity at the single photon scale~\cite{reiserer2014,volz2012, dayan2008, loo2012,bose2012}. Many systems have been explored as quantum nodes in the last decade and have individually shown interesting and highly complementary properties. For instance, defects in diamonds and single atoms have shown the possibility to store the quantum information on the millisecond time scale~\cite{ korber2018}, allowing for the deployment of quantum memory-based quantum networks~\cite{briegel1998}, while semiconductor quantum dots can generate  indistinguishable photons at a high rate  in an scalable way~\cite{ding2016,somaschi2016,senellart2017}. Versatile quantum network architectures would highly benefit from the combination of all these properties, combining various atomic systems as quantum nodes. An important challenge however is the natural bandwidth mismatch of these atomic systems, \Loic{inherent to their different spectral widths and spontaneous emission times.} In this regard, the development of bandwidth tunable atom-photon interfaces is highly desirable.

An efficient atom-photon interface can be obtained by engineering the electromagnetic environment of an atom, placing the quantum emitter in a cavity~\cite{ auffeves2007}. The ideal situation is obtained when the incident light is efficiently injected and collected into and from the cavity~\cite{Gardiner1985} and when the spontaneous emission of the quantum emitter in other modes than the cavity mode is negligible. Systems with good overall efficiencies are obtained with natural atoms~\cite{shomroni2014,reiserer2014, tiecke2014} or semiconductor quantum dots in optical cavities~\cite{giesz2016,bennett2016,snijders2016} as well as superconducting quantum bits in microwave cavities \cite{astafiev2010,ho2011}. In each case, the saturation of the atomic transition at the level of the single photon has been used to demonstrate photon blockade~\cite{shomroni2014,loo2012,bose2012,tiecke2014}, photon-Fock state filtering~\cite{dayan2008,bennett2016, snijders2016,santis2017} as well as photon-photon gates~\cite{shomroni2014,hacker2016}. 

In this work, we propose an interface based on a two-atom system coupled to a single cavity mode and exploit  superradiance and subradiance phenomena to obtain a bandwidth tunable atom-photon interface. In the following, we derive our study in the case of two semiconductor quantum dots (QDs) coupled to a microcavity, a system that has shown state-of-the-art single photon emission~\cite{senellart2017} as well as genuine single-photon non-linearities~\cite{bennett2016,snijders2016,santis2017}. Moreover, this system allows for direct dipole-dipole interaction by using well established growth techniques to vertically stack QDs with a  control of the inter-dot distance at the nanoscale~\cite{Masumoto2002}. Finally, doping of the structure and bias application  allows to tune two QD resonances through the confined Stark effect~\cite{bennett2010}. 

We study the influence of direct dipole-dipole interaction on the collective states of the quantum dots. By an effective Hamiltonian approach, we show that the collective states are robust to detuning only when there is a direct dipole-dipole interaction between the QDs.
We show that in the weak light-matter coupling regime, and with slight detuning in energy of the two atomic systems, both the superradiant and subradiant states resulting from the dipole-dipole coupling can independently be probed, allowing to separately take advantage of the different behaviors of these states. 
In this case, it is possible to \Loic{widely tune the bandwidth} of the –now visible– subradiant state by controlling the frequency mismatch of the QDs. This also allows controlling the single-photon nonlinearity of the system, in particular its nonlinearity threshold and the bandwidth of the photon blockade effect, over orders of magnitude.

The outline of the paper is as follows. In Sec. II we present the theoretical model and the dipole-dipole coupling rate. In Sec. III we study the eigenstates when varying the detuning between the QDs and the direct dipole-dipole coupling. In Sec. IV, the non-linear behavior and the bandwidth of these collective subradiant and superradiant states is analyzed. Sec. V is devoted to the study of the photon blockade effect.
\section{The system}

\begin{figure}
	\includegraphics[width=\columnwidth]{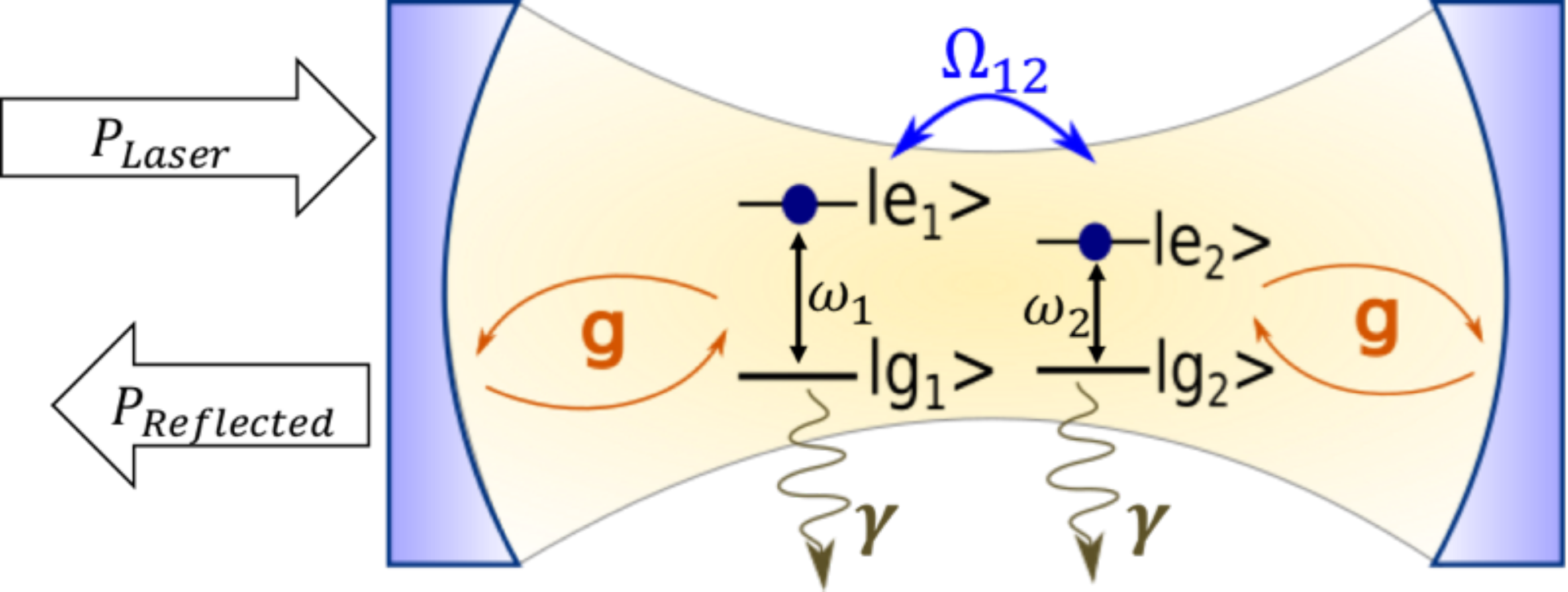}
	\caption{\label{fig:cavity} The CW laser of power $P_{Laser}$ pumps the cavity mode and we measure the reflected power $P_{Reflected}$. Both QDs interact equally with the cavity mode at a rate $g$. The varying parameters are the frequency detuning of the QDs $\Delta_{12}=\frac{\omega_{1}-\omega_{2}}{2}$ and the direct dipole-dipole coupling with a rate $\Omega_{12}$. $\ket{e_{i}}$ and $\ket{g_{i}}$ are the excited and ground state of the i-th QD respectively. The spontaneous decay rates $\gamma$ into the environment can change depending on the direct coupling.}
\end{figure}

The system consists of two quantum dots (QD) coupled with a single mode of an optical solid state cavity as shown in figure~\ref{fig:cavity}. It can be seen as two interacting subsystems which are on the one hand, the two QDs, and on the other hand, the cavity.
The quantum dots are described as two level systems (TLS),  with their ground state $\ket{g_{i}}$ and their excited state $\ket{e_{i}}$ with $i=1,2$ and a transition energy  $E_{\ket{e_{i}}}-E_{\ket{g_{i}}}=\hbar\omega_{i}$. $\gamma$ is the free-space spontaneous emission rate into the environment taken equal for both QDs. The two parameters that control the system are the detuning of the QDs $\Delta_{12}=\frac{\omega_{1}-\omega_{2}}{2}$ and the direct dipole-dipole coupling~\cite{Forster1959} of the QDs. The dipole-dipole coupling is characterized by its energy exchange rate $\Omega_{12}$ and by the modification of the spontaneous emission rates by an amount equal to $\pm \gamma_{12}$~\cite{Ficek2002}.
The QDs are both coupled to the fundamental mode of a Fabry-Perot cavity, whose frequency is $\omega_{c}$, with a rate $g$ which is taken equal for both QDs. The cavity damping rate is $\kappa$ and accounts for radiative losses of the cavity. It can be decomposed as $\kappa=\kappa_{\text{left}}+\kappa_{\text{right}}+\kappa_{\text{other}}$. The first two terms are the damping rates of photons escaping through the cavity mirrors while the last term is the unwanted photon loss in other directions. The desired critical coupling is obtained for $\kappa_{\text{left}}/\kappa=50\text{\%}$ but for simplicity we consider here a symmetric cavity, $\kappa_{\text{left}}=\kappa_{\text{right}}$ and we neglect other dampings $\kappa_{\text{other}}=0$.
The cavity allows to have light-matter interaction at the single photon level and to better collect the emitted light through a partially reflecting mirror~\cite{giesz2016,auffeves2007,tiecke2014}.
The whole system is probed by a continuous wave (CW) incident laser of power $P_{laser}$ pumping the cavity mode by one of the mirrors. The nonlinear behavior of the system is studied by measuring the reflectivity $\frac{P_{Reflected}}{P_{Laser}}$ as a function of the incident power. 
The reflectivity changes due to saturation of the QDs~\cite{loo2012}. The saturation of a single two-level system inside a cavity is described in the steady state by the critical photon number, which corresponds to the average photon number needed inside the cavity to saturate the QD. Without pure dephasing processes of the QDs the critical photon number is given by~\cite{Lanco2015,Armen2006,kimble1994}:
\begin{equation} \label{eq:critical_photon}
n_{c_{0}}=\frac{\gamma^{2}}{8g^{2}}.
\end{equation}
Thus the nonlinear behavior of the system can be modified by changing either the coupling strength or the spontaneous emission rate of the TLS. 
\subsection{Model}
The system is modeled by the driven Tavis-Cummings model~\cite{Tavis1968} for two TLS coupled to a single mode of a cavity.
The Hamiltonian describing the driven cavity containing the two coupled QDs is written in the frame rotating at the frequency of the laser and using the rotating-wave approximation we get with $\hbar=1$~\cite{Lehmberg1970}:
\begin{align}\label{eq:hamiltonian}
H&=(\omega_{1}-\omega_{L})\sigma_{1}^{+}\sigma_{1} + (\omega_{2}-\omega_{L})\sigma_{2}^{+}\sigma_{2} + 
(\omega_{c}-\omega_{L})a^{\dagger}a\nonumber\\
&+i\sqrt{2}g\left(a^{\dagger}\frac{(\sigma_{1}+
\sigma_{2})}{\sqrt{2}}-a\frac{(\sigma_{1}^{+}+\sigma_{2}^{+})}{\sqrt{2}}\right)\nonumber\\
&+\Omega_{12}\left(\sigma_{1}^{+}\sigma_{2}+
\sigma_{1}\sigma_{2}^{+}\right)-i\mathcal{E}_{p} (a^{\dagger}-a)
\end{align}
where $\omega_{i}$, $\omega_{c}$ and $\omega_{L}$ are the frequencies of the i-th  QD, the cavity mode and the laser respectively, $\sigma_{i}=|g_{i}\rangle\langle e_{i}|$ is the lowering operator for the i-th QD, $a$ is the cavity field annihilation operator, $g$ is the coupling between one QD and the cavity mode taken to be equal for both QDs and $\mathcal{E}_{p}$ is the field amplitude of the laser coupled to the cavity mode: $\mathcal{E}_{p}=\sqrt{\frac{\kappa}{2}P_{laser}/\hbar\omega_{L}}$. The dipole-dipole coupling rate $\Omega_{12}$ is written for two parallel dipoles in the dipolar approximation~\cite{Andrews2004,Lehmberg1970,Ficek2002}:
\begin{equation}
	\Omega_{12}=\Re{\gamma\,F(kd)}
\end{equation}
with:
\begin{equation} \label{eq:tenseurgreen}
F(x)=-\frac{3}{4}e^{ix}\left[\frac{1}{x}+\frac{i}{(x)^{2}}-\frac{1}{(x)^{3}}
\right] \, ,
\end{equation}
$d$ the distance separating the two QDs and $k=\frac{2\pi}{\lambda}$ the wavevector in the bulk material. For InGaAs QDs in GaAs it is close to $\lambda=930\,\text{nm}/n_{GaAs}$, with $n_{GaAs}=3.6$~\cite{E1985}.
When the two quantum dots are close to each other, $kd\ll 1$, so that:
\begin{equation}\label{eq:omega12}
\Omega_{12}\simeq\gamma\frac{3}{4\left(kd\right)^{3}}\, .
\end{equation}
The open system evolution is best described using the density matrix to take into account the interaction with the environment. The coherent evolution is described by the Hamiltonian~\cite{Bonifacio1971}:
\begin{equation}
	\left\{\dot{\rho}\right\}_{coh}=i\comm{\rho}{H}.
\end{equation}
The incoherent coupling to the environment is described by:
\begin{equation}
	\left\{\dot{\rho}\right\}_{incoh}=\kappa\mathcal{L}(a)+
	\sum_{i}\left(\gamma\mathcal{L}(\sigma_{i})+\gamma^{*}
	\mathcal{L}(\sigma^{+}_{i}\sigma_{i})
	\right)
\end{equation}
where for a given operator $\hat{o}$, $\mathcal{L}(\hat{o})=\hat{o}\rho \hat{o}^{\dagger}-\frac{1}{2}\hat{o}^{\dagger} \hat{o}\rho-\frac{1}{2}\rho \hat{o}^{\dagger} \hat{o}$. The first term describes the leak of a cavity excitation through the mirrors with a rate $\kappa$. The next term describes free-space spontaneous emission of the i-th QD with a rate $\gamma$, i.e. the emission of the quantum dot into electromagnetic modes other than the cavity mode (the so called environment or leaky modes). The final term describes pure dephasing at a rate $\gamma^{*}$. 
In practice collective effects are observable for $\gamma^{*}$ smaller than $\frac{4g^2}{\kappa}$. State of the art systems~\cite{giesz2016,snijders2016} have obtained negligible pure dephasing and they are not taken into account in the rest of this article, i.e. we set $\gamma^{*}=0$. 
Dipole-dipole interaction is responsible for a modification in the free-space spontaneous emission of the QDs. It corresponds to a new term in the incoherent part of the equation which is written~\cite{Lehmberg1970,Hettich2002}:
\begin{equation}
	\gamma_{12}\,\sum_{i\neq j}\left(\sigma_{i}\rho \sigma^{+}_{j}-\frac{1}{2}\sigma_{j}^{+}\sigma_{i}\rho-\frac{1}{2}\rho\sigma_{j}^{+}\sigma_{i}\right)
\end{equation}
with 
\begin{equation}
\gamma_{12}=-\frac{1}{2}\Im{\gamma\,F(kd)}
\end{equation}
with $F(x)$ given in Eq.~\ref{eq:tenseurgreen}. Again in the small distance approximation, $kd\ll 1$:
\begin{equation}\label{eq:gamma12}
	\gamma_{12}\simeq\gamma\,.
\end{equation} 
When adding all the incoherent terms together and after rearrangement we obtain:
\begin{align}\label{eq:lindbladians}
	\left\{\dot{\rho}\right\}_{incoh}&=\kappa\mathcal{L}(a)+(\gamma+\gamma_{12})\,\mathcal{L}\left(\frac{\sigma_{1}+\sigma_{2}}{\sqrt{2}}\right)\nonumber\\
	&+(\gamma-\gamma_{12})\,\mathcal{L}\left(\frac{\sigma_{1}-\sigma_{2}}{\sqrt{2}}\right).
\end{align}
It can be seen that dipole-dipole interaction modifies spontaneous emission of the coupled QDs into the leaky modes: it increases the emission rate by $\gamma_{12}$ for in phase emission (emission of the symmetric state) and it decreases the rate by $\gamma_{12}$ for out of phase emission (emission of the antisymmetric state). It can be noted that both the coherent energy exchange rate $\Omega_{12}$ and the modification of the incoherent spontaneous emission by $\gamma_{12}$ due to dipole-dipole coupling, are described by the same function $F$ and that for large distances between the QDs they both tend to 0. 
This limiting case, i.e. $\Omega_{12}=0$ and $\gamma_{12}=0$ which corresponds to isolated QDs, will be taken as a reference in what follows, and will be referred as $\Omega_{12}=0$.
In this article we consider self assembled InGaAs QDs that can be vertically stacked with small separation distances $d\ll\lambda$~\cite{Michler2009}. In this case $\gamma_{12}\simeq\gamma$ and the spontaneous emission of the fully antisymmetric state is almost suppressed while the emission of the symmetric state is doubled. Finally the master equation can be cast in the form:
\begin{align}\label{eq:equation-maitresse}
\dot{\rho}&=i\comm{\rho}{H}+\kappa\mathcal{L}(a)+(\gamma+\gamma_{12})\,\mathcal{L}\left(\frac{\sigma_{1}+\sigma_{2}}{\sqrt{2}}\right)\nonumber\\
&+(\gamma-\gamma_{12})\,\mathcal{L}\left(\frac{\sigma_{1}-\sigma_{2}}{\sqrt{2}}\right).
\end{align}
We use parameter values that correspond to typical experiments of self-assembled InGaAs QDs in micropillar cavities~\cite{somaschi2016} but that are also valid for other solid-state CQED experiments with two-level systems like color centers~\cite{Sipahigil2014} or QDs in photonic crystal cavities~\cite{Laucht2010}. The values are $\left\{g,\kappa,\gamma\right\}=\left\{ 20, 200, 0.6 \right\}\,\si{\micro\electronvolt}$, corresponding to a system of high cooperativity~\cite{Reiserer2015} (thus high Purcell factor) $C=\frac{2g^2}{\kappa\gamma}=7$, but still in the weak coupling or bad-cavity regime. For a negligible pure dephasing, the cooperativity is simply equal to half of the Purcell factor which is given by the cavity-enhanced decay rate of the QD, 
\begin{equation}\label{eq:Gamma_purcell}
	\Gamma_{0}=\frac{4g^{2}}{\kappa} \, ,
\end{equation} over the decay rate into free space $\gamma$. So in the bad cavity high cooperativity regime, called also the Purcell regime, the quantum dots interact mostly with the cavity mode and consequently can efficiently be probed. In this regime there is no vacuum Rabi splitting and the states are mainly excitonic for the QDs and photonic for the cavity mode.
In this system, the light-matter interaction depends on the direct coupling $\Omega_{12}$ between the QDs and also on the detuning $\Delta_{12}$ between them. To identify the effects of these two parameters, we consider four separate cases: first the simple case of two identical and independent QDs. We will then separately study the effect of detuning and direct dipole-dipole interaction. And finally we will look at the system with both detuning and direct coupling of the QDs. 

\section{System eigenstates}\label{sec:eigenstates}
\subsection{Identical QDs: $\bm{\Delta_{12}=0}$ and $\bm{\Omega_{12}=0}$}
\begin{figure}
	\includegraphics[width=\columnwidth]{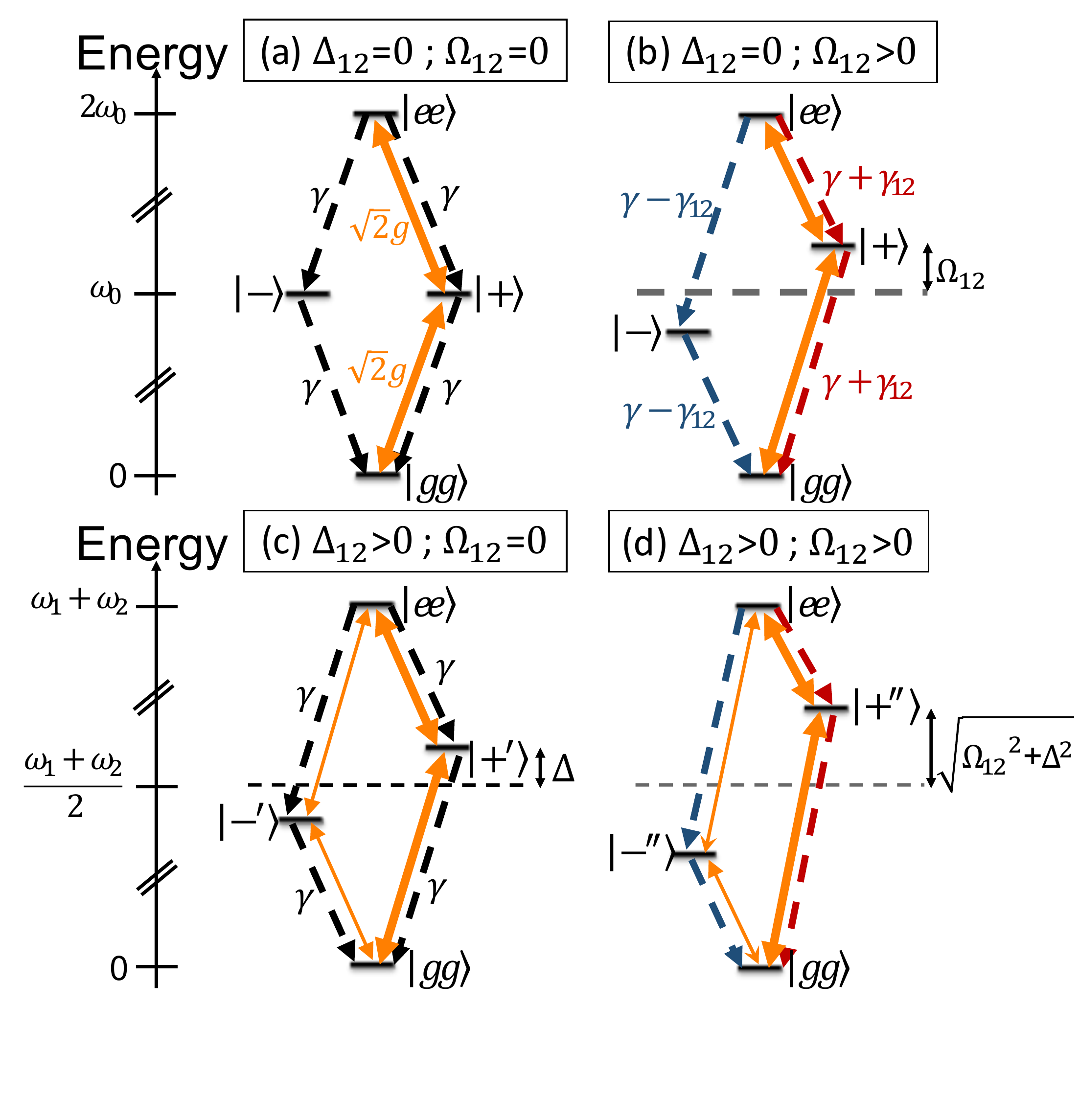}
	\caption{\label{fig:diagramme_energie}Energy diagram for the QDs. $\ket{ee}$ and $\ket{gg}$ represent the states with both QDs excited or in the ground state. The other states are entangled states and are described in the main text. The dotted arrows represent spontaneous decays and the double arrows depict coherent coupling of levels through the cavity mode. \textbf{(a)} $\mathbf{\Omega_{12}=0}$, $\mathbf{\Delta_{12}=0}$, only the symmetric state $\ket{+}$ couples to the cavity by absorbing or emitting a photon with a rate $\sqrt{2}g$. The antisymmetric state $\ket{-}$ is a dark state and doesn't interact with the cavity. The spontaneous decay is the same for all states and is equal to the single QD decay rate $\gamma$.
	\textbf{(b)} $\mathbf{\Omega_{12}>0}$, $\mathbf{\Delta_{12}=0}$, QDs are dipole-dipole coupled. The entangled eigenstates are still completely symmetric and antisymmetric but they are now energy split by $\pm\Omega_{12}$ and the spontaneous decay rates are modified by $\pm\gamma_{12}$ respectively.
	\textbf{(c)} $\mathbf{\Omega_{12}=0}$, $\mathbf{\Delta_{12}>0}$, QDs are detuned,  the new entangled eigenstates $\ket{-'}$ and $\ket{+'}$ are a superposition of the previous totally symmetric and antisymmetric states. $\ket{-'}$ now couples to the cavity mode. There is no dipole-dipole coupling so the spontaneous decay rates are not modified and are still equal to $\gamma$ for all the states.
	\textbf{(d)} $\mathbf{\Omega_{12}>0}$, $\mathbf{\Delta_{12}>0}$ Detuned and dipole-dipole coupled QDs: the eigenstates are energy split by $\sqrt{\Omega_{12}^2+\Delta_{12}^{2}}$, $\ket{+''}$ and $\ket{-''}$ are both coupled to the cavity and the spontaneous decay rates are modified.}
\end{figure}

First we consider the case with identical QDs ($\Delta_{12}=0$), that are not directly coupled to each other, ($\Omega_{12}=0$).
The two QDs couple only via the cavity mode. This is the textbook Tavis-Cummings case~\cite{Armen2006,Tavis1968}, even if here the system is in the Purcell regime~\cite{Casabone2015}.
The four eigenstates of the QDs consist of the antisymmetric singlet $\ket{-}=\frac{\ket{e,g}-\ket{g,e}}{\sqrt{2}}$ and the symmetric triplet $\left\{\ket{gg} ;\ket{+}=\frac{\ket{e,g}+\ket{g,e}}{\sqrt{2}}; \ket{ee}\right\}$ as shown in figure~\ref{fig:diagramme_energie}(a).
The antisymmetric state $\ket{-}$ does not couple to the cavity, it is a dark state. 
The coupling of the symmetric state $\ket{+}$ with the cavity mode is enhanced: it now scales as $\sqrt{2}g$ due to the coherent coupling with the cavity mode.
Concerning free-space spontaneous emission outside the cavity mode, since the two QDs are not directly coupled, the two eigenstates $\ket{+}$ and $\ket{-}$ spontaneously decay to the ground state with the same rate $\gamma$ of a single QD. This amounts to consider that $kd\gg 1$ which means $\gamma_{12}\simeq 0$.
Finally, without dipole-dipole coupling or detuning ($\Omega_{12}=0$), the energy states are not split so the transition $\ket{gg}\leftrightarrow \ket{+}$ is resonant with $\ket{+}\leftrightarrow \ket{ee}$ as shown in the diagram. The system can then absorb two photons with the same frequency so there is no single photon blockade effect.
\subsection{Dipole-dipole coupling only, $\bm{\Delta_{12}=0}$ and $\bm{\Omega_{12}\neq0}$ }
We now take dipole-dipole interaction into account by setting the distance of the two QDs to $d=10\,\text{nm}$.
The eigenstates of the QDs are still completely symmetric and antisymmetric but the energies are split and the spontaneous decay rates in free space change as shown in \textbf{figure}~\ref{fig:diagramme_energie}(b).
For stacked QDs the dipoles are parallel and the symmetric state $\ket{+}$ is blue shifted while the antisymmetric state $\ket{-}$ is red shifted by $\Omega_{12}$ respectively. 
The free-space decay rates are modified as specified by Eq.~\ref{eq:equation-maitresse}:
spontaneous decay through the symmetric branch $\ket{ee}\rightarrow \ket{+} \rightarrow \ket{gg}$ is enhanced and it is now equal to $\gamma+\gamma_{12}\simeq 2\gamma$. The state $\ket{+}$ is called superradiant. 
The free-space decay rate is decreased for the antisymmetric branch $\ket{ee}\rightarrow \ket{-} \rightarrow \ket{gg}$ and it is equal to $\gamma-\gamma_{12}\simeq 0$. The state $\ket{-}$ is called subradiant. 
It is completely antisymmetric and stays dark as it does not couple to the cavity mode and cannot be excited. The superradiant state is  completely symmetric so it couples as $\sqrt{2}g$ with the cavity mode. But since the transitions $\ket{gg}\leftrightarrow \ket{+}$ and $\ket{+}\leftrightarrow \ket{ee}$ are not resonant anymore, the superradiant state $\ket{+}$ behaves now as an effective two-level system in the sense that at most one photon at the frequency of the transition $\ket{gg}\leftrightarrow \ket{+}$ can be simultaneously absorbed. One can then use Eq.~\ref{eq:critical_photon}, and obtain the critical photon number to saturate this transition:
\begin{equation}\label{eq:critical_photon_sup}
	n_{c_{\ket{+}}}=\frac{(\gamma+\gamma_{12})	^{2}}{8(\sqrt{2}g)	^2}\simeq \frac{\gamma^{2}}{4g	^2}  =2n_{c_{0}}\, ,
\end{equation}
with $\gamma_{12}\simeq\gamma$.
Twice as much intensity is needed to saturate the superradiant state compared to a single QD due to its larger spontaneous emission rate into the leaky modes.

\subsection{Detuned QDs with no direct coupling, $\bm{\Delta_{12}\neq0}$ and $\bm{\Omega_{12}=0}$}
We now study the influence of detuning by taking non identical QDs with different bare frequencies so that $\Delta_{12}\neq0$. We set for the moment $\Omega_{12}=\gamma_{12}=0$, i.e. $kd\gg 1$, the dipole-dipole interaction will be added in the next paragraph. The energy diagram of this case is sketched in figure~\ref{fig:diagramme_energie}(c). The state $\ket{gg}$ and $\ket{ee}$ stay unchanged but the new entangled states $\ket{+'}$ and $\ket{-'}$ are now a mixture of the previous symmetric and antisymmetric states $\ket{+}$ and $\ket{-}$. 
This case has been studied theoretically and experimentally in the strong coupling regime~\cite{Albert2013,Averkiev2009,Laucht2010,Radulaski2017}. The interesting feature is the appearance of a new peak in the scattering spectrum of the system when the QDs are detuned. This new peak corresponds to the antisymmetric state which is totally dark when the QDs are resonant $\Delta_{12}=0$ but that starts to couple to the cavity mode once a detuning between the QDs is introduced ($\Delta_{12}\neq0$)~\cite{DeLeseleuc2017}. To our knowledge this system hasn't been studied in the Purcell regime. In this regime we can focus on the new entangled states of the QDs which are written for $g\ll\kappa$:
\begin{align} \label{eq:detuned_modes}
&\ket{-'} \simeq A \ket{+} + B \ket{-}  \nonumber\\
&\ket{+'} \simeq B \ket{+} - A\ket{-} ,
\end{align}
where A and B are coefficients that depend on $\Delta_{12}$. Their evolution in the linear (low power) regime is plotted in figure~\ref{fig:AB_fx_delta}. It has been calculated using the effective Hamiltonian approach, see appendix \ref{app:effective_hamiltonian}. 
\begin{figure}
	\includegraphics[width=\columnwidth]{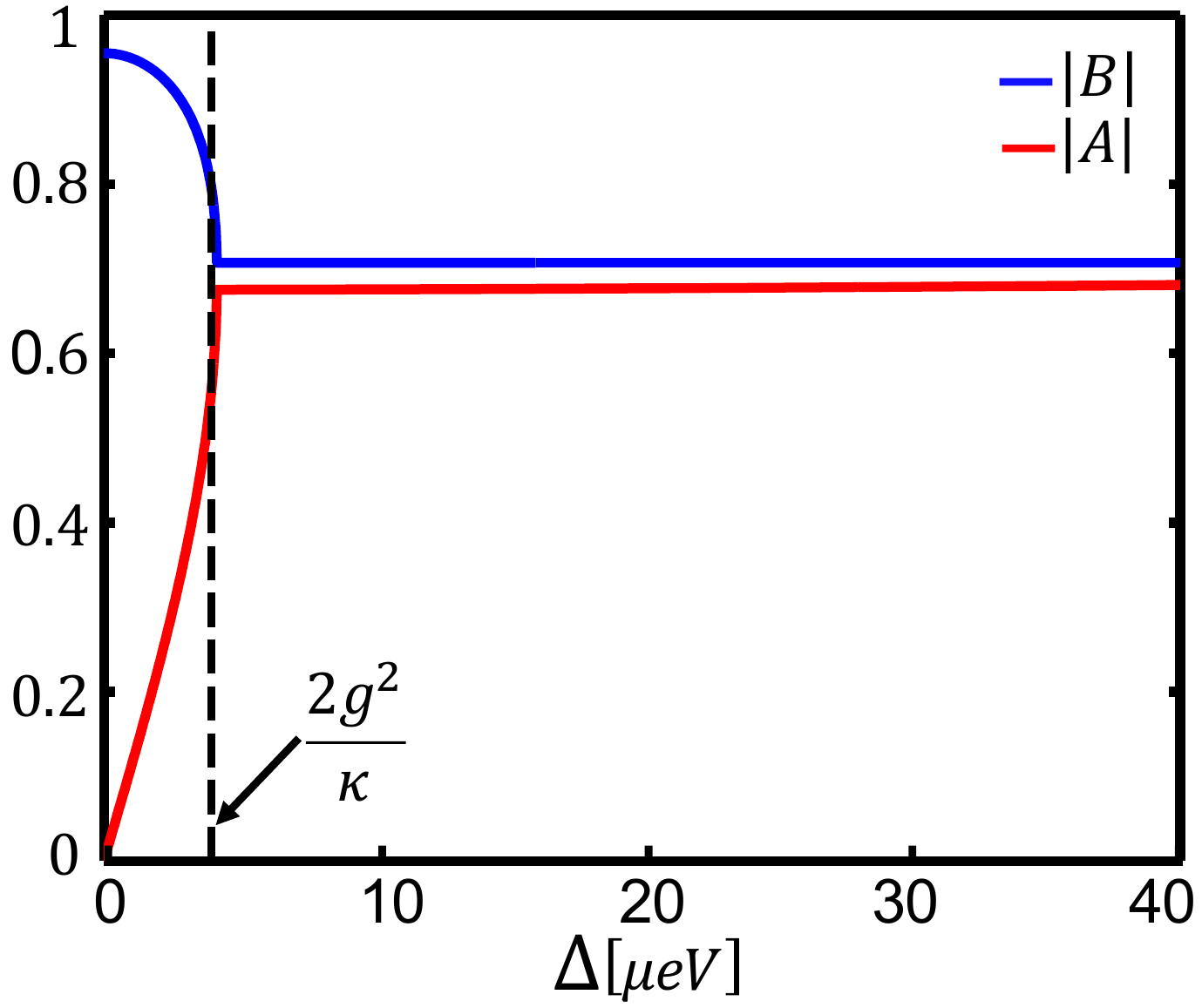}
	\caption{\label{fig:AB_fx_delta}Modulus of the decomposition coefficients A (lower curve) and B (upper curve) for the entangled states $\ket{+'}$ and $\ket{-'}$ of Eq.~\ref{eq:detuned_modes} as a function of detuning between the two QDs. Parameters are \{$g$,$\gamma$,$\kappa$\}=\{20, 0.6, 200\} $\si{\micro\electronvolt}$. The vertical line indicates detuning $\Delta_{12}=\Gamma_{0}$.}
\end{figure}
Two distinct regimes appear depending on the value of detuning. For $\Delta_{12}<\frac{2g^{2}}{\kappa}$, $\abs{B}$ is higher than $\abs{A}$, so following Eq.~\ref{eq:detuned_modes} it means that the eigenstates $\ket{+'}$ and $\ket{-'}$ have a high symmetric and antisymmetric component respectively. Indeed $B=\braket{+}{+'}=\braket{-}{-'}$. The two QDs are coupled through the cavity mode and the collective states appear. For $\Delta_{12}>\frac{2g^{2}}{\kappa}$, $\abs{A}$ and $\abs{B}$ are each equal to $\frac{1}{\sqrt{2}}$, the eigenstates $\ket{+'}$ and $\ket{-'}$ have lost their respective symmetric and antisymmetric character and now behave as independent QDs.
The limiting value separating the two different regimes is 
\begin{equation}\label{eq:couplage_cav_limite}
\Delta_{12}=\frac{2g^{2}}{\kappa}=\frac{\Gamma_{0}}{2}\, ,
\end{equation}
the cavity-enhanced decay rate of a single QD of Eq.~\ref{eq:Gamma_purcell} over 2. 
The two different regimes arise due to the fact that when the total detuning $2\Delta_{12}$ is shorter than the decay rate in the cavity $\Gamma_{0}$ (i.e. beating period between the two QDs longer than the Purcell-enhanced decay time) the two QDs can effectively couple before they desynchronize. Whereas when the total detuning $2\Delta_{12}$ is larger than the decay rate in the cavity $\Gamma_{0}$ (i.e. beating period between the two QDs shorter than the Purcell-enhanced decay time), desynchronization between the QDs suppresses the symmetric and antisymmetric character of the states. Thus, in the Purcell regime without direct coupling, it is not possible to probe the collective states independently (by detuning them enough) without suppressing their collective behavior.
In the next paragraph we will see that introducing a direct interaction between the QDs allows recovering the symmetric and antisymmetric character and thus the collective behavior of the coupled states even for $\Delta_{12}>\frac{\Gamma_{0}}{2}$.
%
%
\subsection{Dipole-dipole coupled AND detuned QDs, $\bm{\Delta_{12}\neq0}$ and $\bm{\Omega_{12}\neq 0}$}
\begin{figure}
	\includegraphics[width=\columnwidth]{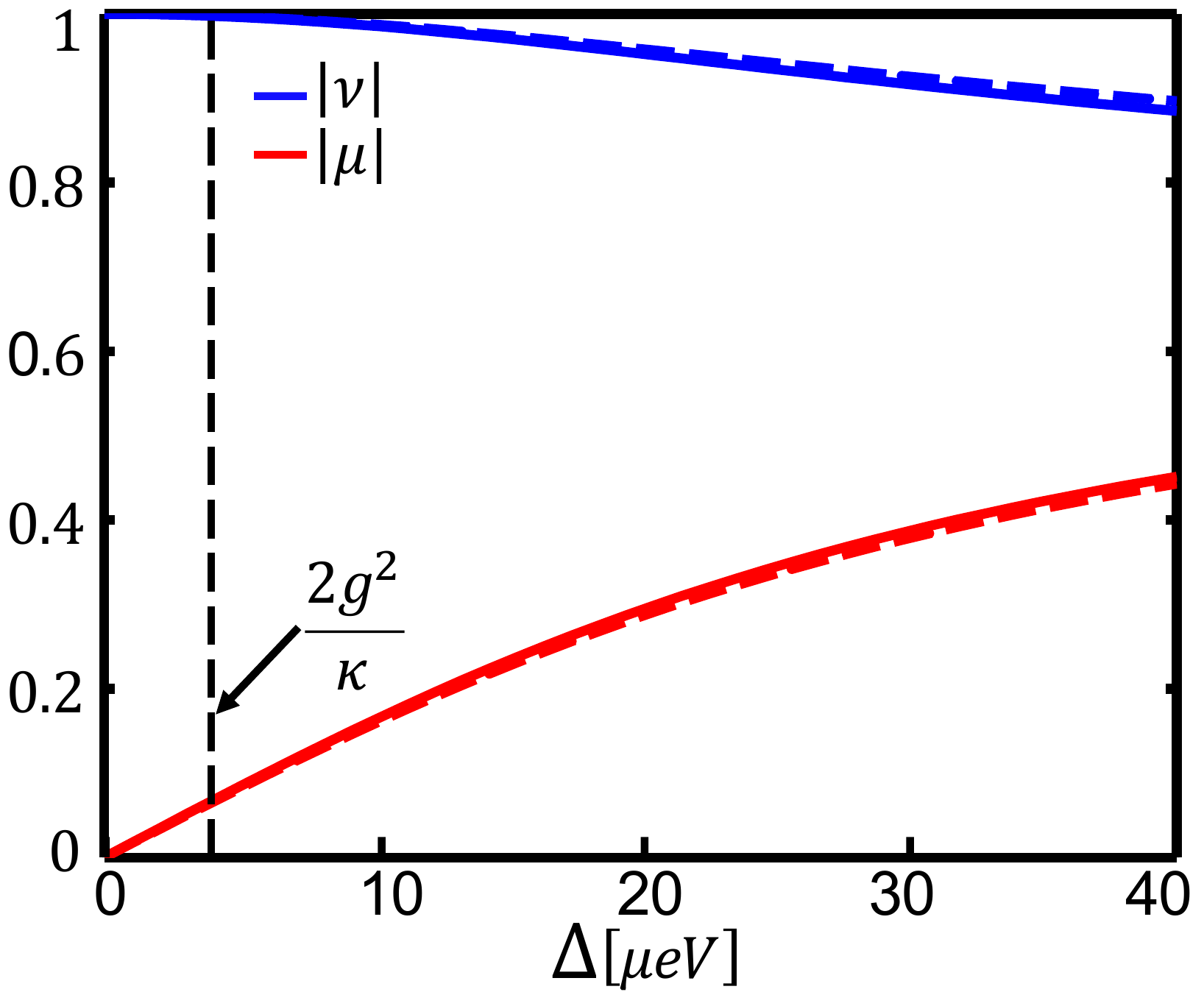}
	\caption{\label{fig:MuNu_fx_delta}Modulus of the decomposition coefficients $\mu$ (lower curves) and $\nu$ (upper curves) for the entangled states $\ket{+''}$ and $\ket{-''}$ of Eq.~\ref{eq:detuned_coupled_modes}. Dashed lines show the analytical results of Eq.~\ref{eq:munu_analytic}. Parameters are the same as figure~\ref{fig:AB_fx_delta} but now $\Omega_{12}=31 \, \si{\micro\electronvolt}$ corresponding to a distance of $d=10\,\text{nm}$.}
\end{figure}

To be able to address separately the superradiant and subradiant states and also tune their spontaneous decay rates, we now consider the case when the two QDs are both detuned and directly coupled. 
Detuning will allow the interaction of the cavity mode with the two entangled states, including the subradiant state, as depicted in figure~\ref{fig:diagramme_energie}(d).
And dipole-dipole interaction will allow to maintain the collective behavior of the states at larger detunings.
The states are now written $\ket{+''}$ and $\ket{-''}$. As before their symmetric and antisymmetric component will depend on the detuning. The decomposition into these states can be written in the same way as before for $g\ll\kappa$ but with different coefficients: 
\begin{align} \label{eq:detuned_coupled_modes}
&\ket{-''}\simeq\mu \ket{+} + \nu \ket{-}  \nonumber\\
&\ket{+''}\simeq\nu \ket{+} - \mu\ket{-} ,
\end{align}
where $\mu$ and $\nu$ are the new coefficients of the decomposition. Their modulus is plotted in figure~\ref{fig:MuNu_fx_delta}. They are calculated numerically by diagonalizing the effective Hamiltonian as before, see appendix~\ref{app:effective_hamiltonian}. An analytical approximation is obtained when considering only the subsystem of the coupled QDs without the cavity mode, we can then write $\mu$ and $\nu$ as function of only the atomic parameters $\Delta_{12}$ and $\Omega_{12}$. We find~\cite{Ficek1986}:
\begin{align} \label{eq:munu_analytic}
\mu&=\frac{\delta}{\sqrt{\delta^{2}+\left(1+\sqrt{1+\delta^{2}}\right)^{2}}} \nonumber \\
\nu&=\frac{1+\sqrt{1+\delta^{2}}}{\sqrt{\delta^{2}+\left(1+\sqrt{1+\delta^{2}}\right)^{2}}}
\end{align}
with $\delta=\frac{\Delta_{12}}{\Omega_{12}}$.
The analytical value, plotted in the same figure, is very close to the numerical solution and this for the whole range of $\Delta_{12}$ considered. This means that in the case of direct coupling in the Purcell regime the collective states depend mostly on the direct coupling rate $\Omega_{12}$. Note that in this limit, $\mu$ and $\nu$ do not depend on $\frac{g^{2}}{\kappa}$. As it can be seen, $\abs{\nu}$ stays close to $1$ for a much larger range of $\Delta_{12}$ than before, which means $\ket{+''}$ and $\ket{-''}$ have a larger symmetric and antisymmetric component respectively over a wider detuning range. So even for large detuning, we will have $\ket{+''}$ behaving mostly as a symmetric state $\ket{+}$ and $\ket{-''}$ mostly as an antisymmetric state $\ket{-}$. Dipole-dipole coupling thus allows to probe the coupled states separately and to take advantage of their different properties even in the bad cavity regime. 
As both entangled states possess a fraction of the completely symmetric state $\ket{+}$, both couple to the cavity mode. These rates are calculated in appendix~\ref{app:cavity_coupling} and we obtain:
\begin{align} \label{eq:coupling''}
g_{\ket{-''}}&= \mu \sqrt{2}g\nonumber\\
g_{\ket{+''}}&= \nu \sqrt{2}g \, .
\end{align}
In addition, due to dipole-dipole interaction, $\ket{+''}$ has an enhanced free-space spontaneous emission and $\ket{-''}$ a reduced free-space spontaneous emission. Assuming $d\ll\lambda$, the decay rate of these states are given by:
\begin{align} \label{eq:spont''}
\gamma_{\ket{-''}}&=\mu^{2}(\gamma+\gamma_{12})\simeq 2\mu^{2}\gamma \nonumber\\
\gamma_{\ket{+''}}&=\nu^{2}(\gamma+\gamma_{12}) \simeq 2\nu^{2}\gamma \, ,
\end{align}
where we have neglected terms in $\gamma-\gamma_{12}\simeq 0$.
Equations~\ref{eq:coupling''} and~\ref{eq:spont''} show that both the coupling to the cavity modes and the spontaneous decay rates now depend on the detuning and on the direct coupling.
Furthermore since the states are energy split by $\pm\sqrt{\Delta_{12}^{2}+\Omega_{12}^{2}}$, the transitions $\ket{gg}\leftrightarrow\ket{-''}$ and $\ket{-''}\leftrightarrow\ket{ee}$ are not resonant and neither are  $\ket{gg}\leftrightarrow\ket{+''}$ and $\ket{+''}\leftrightarrow\ket{ee}$. So we can consider the states $\ket{-''}$ and $\ket{+''}$ as effective TLS and we can then use the same expression for the critical photon number. When replacing $g$ and $\gamma$ by their expression~\ref{eq:coupling''} and~\ref{eq:spont''} we get:
\begin{align} \label{eq:critical_photon''}
	n_{c_{\ket{-''}}}&=\frac{(\mu^{2}(\gamma+\gamma_{12}))^{2}}{8(\mu \sqrt{2}g)^{2}}
	\simeq \frac{(2\mu^{2}\gamma)^{2}}{8(\mu \sqrt{2}g)^{2}}  
	= 2\mu^{2} n_{c_{0}}\\
	n_{c_{\ket{+''}}}&=\frac{(\nu^{2}(\gamma+\gamma_{12}))^{2}}{8(\nu \sqrt{2}g)^{2}} 
	\simeq \frac{(2\nu^{2}\gamma)^{2}}{8(\nu \sqrt{2}g)^{2}} 
	= 2\nu^{2} n_{c_{0}} \, .
\end{align}
Since $\mu\ll1$ the nonlinear behavior of the subradiant state $\ket{-''}$ is enhanced. For the parameters considered here and $\Delta_{12}=20~\si{\micro\electronvolt}$, we obtain:
$n_{c_{\ket{-''}}}= 0.16 \, n_{c_{0}}$, so the subradiant state should saturate for an incident pump laser intensity an order of magnitude smaller than the case of a single QD.
Instead, the superradiant state $\ket{+''}$ is more robust to saturation $n_{c_{\ket{+''}}}= 1.8 \, n_{c_{0}}$.

\section{Tunable bandwidth and nonlinearity}
\begin{figure*}
	\includegraphics[width=1\textwidth]{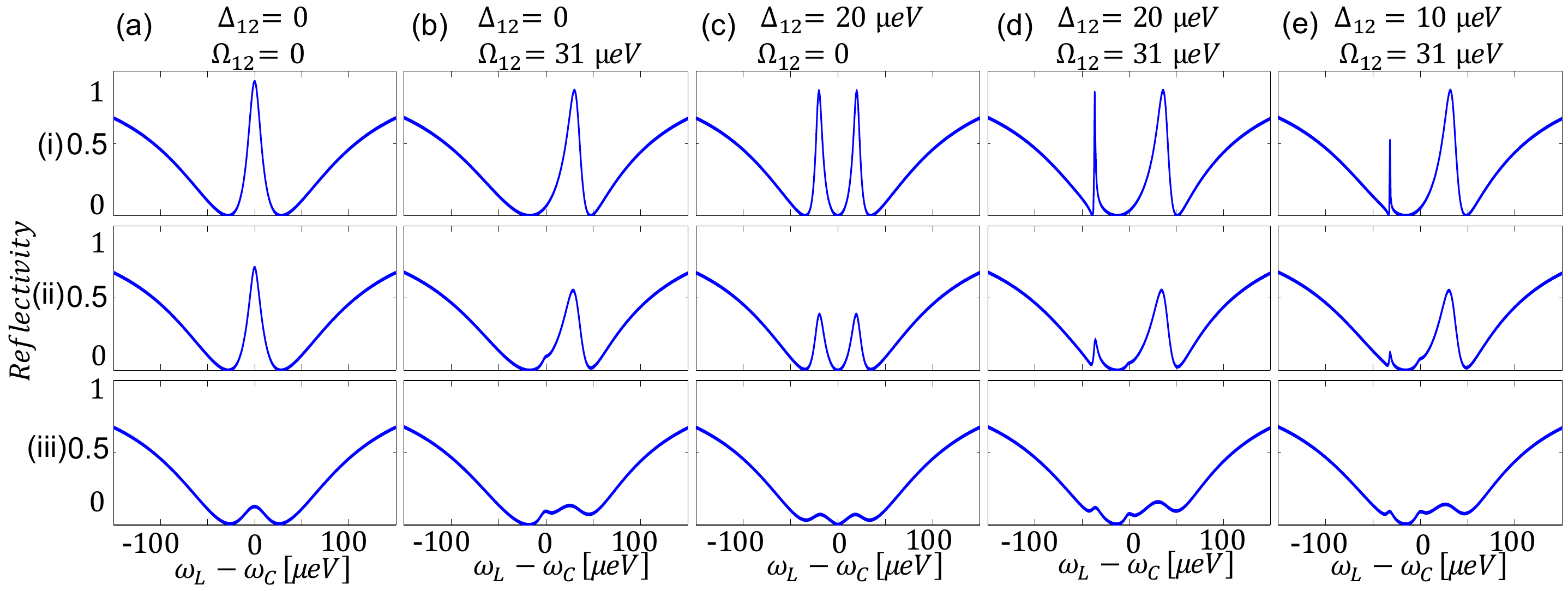}
	\caption{\label{fig:spectres_cw}Reflectivity of the cavity containing two QDs. The columns represent: (a) QDs in resonance and without dipole-dipole coupling. (b) and (c) are the cases with dipole-dipole coupling and detuning respectively. (d) The two QDs are both detuned and dipole-dipole coupled. (e) Same as (d) but with smaller detuning. The three rows show the same spectra but for increasing incident laser power of (i) $1pW$ (ii) $1nW$ and (iii) $10nW$. Parameters are $\left\{g,\kappa,\gamma\right\}=\left\{ 20, 200, 0.6 \right\}\,\si{\micro\electronvolt}$.}
\end{figure*}

The previous discussion has shown that the combination of detuning and direct coupling allows to obtain subradiant and superradiant states with different coupling to the cavity mode and saturation behavior The resulting modification of the linewidth and of the non-linear behavior of the system will now be explored by calculating the power dependent reflectivity spectra.
All calculations have been performed using the MATLAB quantum optics toolbox which numerically solves the master equation for the density matrix~\cite{Tan1999}.
The reflectivity spectra of the system in the four different cases considered before are shown in figure~\ref{fig:spectres_cw}. 
In all four cases, the Fabry-Perot cavity mode appears as a dip in reflectivity while the different peaks indicate states involving the QDs. The plots in each of the four columns represent a different set of parameters corresponding to the four different cases of figure~\ref{fig:diagramme_energie}. The rows represent the same experiments but with increasing pump power. First row (i) is for $P_{laser}=1 pW$, the second row (ii) for $P_{laser}=1 nW$ and the third row (iii) for $P_{laser}=10 nW$. As the states saturate they interact less with the cavity mode and their peak's height is reduced until the empty cavity behavior is retrieved. 

\textbf{(a) $\bm{\Delta_{12}=0}$ and $\bm{\Omega_{12}=0}$}. In the first column of the figure, the QDs are resonant and not directly coupled, i.e. $\Delta_{12}=0$ and $\Omega_{12}=0$, only the symmetric state $\ket{+}$ is coupled to the cavity mode and appears as a large peak at $\omega_{c}$ as shown in figure~\ref{fig:spectres_cw}(a). The peak width is equal to the cavity enhanced spontaneous emission rate of this state as written in Eq.~\ref{eq:Gamma_purcell}, except that now the coupling to the cavity mode is $\sqrt{{2}g}$ so 
\begin{equation}
\Gamma_{\ket{+}}=\frac{8g^{2}}{\kappa}\, .
\end{equation}
The other entangled state, $\ket{-}$, is dark and does not appear in the spectrum, $\Gamma_{\ket{-}}=0 $. When increasing the pump power, the $\ket{+}$ state starts to saturate, the peak's height is reduced. Nevertheless since the transition $\ket{gg}\leftrightarrow \ket{+}$ is resonant with $\ket{+}\leftrightarrow \ket{ee}$, the absorption of a second photon at the same frequency can take place preventing saturation at the single photon level. Thus the peak corresponding to the $\ket{+}$ state stays visible even at high power.

\textbf{(b) $\bm{\Delta_{12}=0}$ and $\bm{\Omega_{12}\neq 0}$}. 
Fig.~\ref{fig:spectres_cw}(b) displays the case of no detuning but with dipole-dipole coupling. The coupling to cavity mode of the two eigenstates remains the same as in the previous case and hence the linewidths are unchanged. The subradiant state is still totally antisymmetric, so it is dark and does not appear in the reflected light whereas the superradiant state has the same width than the symmetric state of figure~\ref{fig:spectres_cw}(a),  but now its energy is shifted by $\Omega_{12}$. The transitions $\ket{gg}\leftrightarrow \ket{+}$ and $\ket{+}\leftrightarrow \ket{ee}$ are not resonant anymore, and the new superradiant state $\ket{+}$ acts now as an effective TLS. When the laser power is increased, this state saturates quicker than in the first column as can be seen when comparing the peak's height.

\textbf{(c) $\bm{\Delta_{12}\neq 0}$ and $\bm{\Omega_{12}=0}$}. In the column Fig.~\ref{fig:spectres_cw}(c), only detuning between the QDs is considered and no direct dipole-dipole interaction. Two peaks appear at the frequencies $\pm\Delta_{12}$ corresponding to the states $\ket{+'}$ and $\ket{-'}$ of Eq.~\ref{eq:detuned_modes}. But as discussed before, without dipole-dipole coupling, the collective states behave as the states of the independent QDs $\ket{e,g}$ and $\ket{g,e}$ as soon as $\Delta_{12}>\frac{2g^{2}}{\kappa}$. \ilan{This is the case for the detuning considered here} and that is why the two peaks that appear are identical and behave as the states of each independent QD $\ket{eg}$ and $\ket{ge}$ interacting independently with the cavity mode. Their linewidth is equal to the linewidth of the single QD: 
\begin{equation}
\Gamma_{\ket{+'}}\simeq\Gamma_{\ket{-'}}\simeq\Gamma_{0} \, .
\end{equation} 
Concerning the non-linear behavior, when pump power is increased they saturate as a single QD would do, and since the critical photon number for a single QD is smaller than for the superradiant state as stated in Eq.~\ref{eq:critical_photon_sup}, they saturate faster than the state $\ket{+}$ of column (b).

\textbf{(d) and (e) $\bm{\Delta_{12}\neq 0}$ and $\bm{\Omega_{12}\neq 0}$}. 
If now we consider both detuned and dipole-dipole coupled QDs as shown in figure~\ref{fig:spectres_cw}(d,e) not only we observe the blue shifted superradiant state $\ket{+''}$ but also the red shifted much narrower antisymmetric state $\ket{-''}$ that now couples to the cavity. Their different linewidth is due to different coupling to the cavity mode: $\ket{-''}$ couples very slightly to the cavity as opposed to $\ket{+''}$ as presented in Eq.~\ref{eq:coupling''}. When replacing their coupling rate to the cavity mode we obtain linewidths equal to
\begin{equation}\label{eq:Gamma+''}
	\Gamma_{\ket{+''}}=\frac{\nu^{2}8g^{2}}{\kappa}
\end{equation} and
\begin{equation}\label{eq:Gamma-''}
	\Gamma_{\ket{-''}}=\frac{\mu^{2}8g^{2}}{\kappa}\, ,
\end{equation}
\ilan{with $\mu$ and $\nu$ given by equation~\ref{eq:munu_analytic}.}
And as expected from the critical photon number of these two states shown in Eq.~\ref{eq:critical_photon''}, the subradiant state $\ket{-''}$ saturates for a much lower incident power. Indeed already in the second row its height is greatly reduced. A remarkable feature of this system is that both the linewidth and the saturation intensity can be tuned by controlling $\Delta_{12}$. This is directly seen when comparing the columns (d) with $\Delta_{12}=20\, \si{\micro\electronvolt}$ and (e) with  $\Delta_{12}=10\, \si{\micro\electronvolt}$: the state $\ket{-''}$ linewidth is reduced and it saturates at lower laser power for smaller detunings.

\begin{figure}
	\includegraphics[width=\columnwidth]{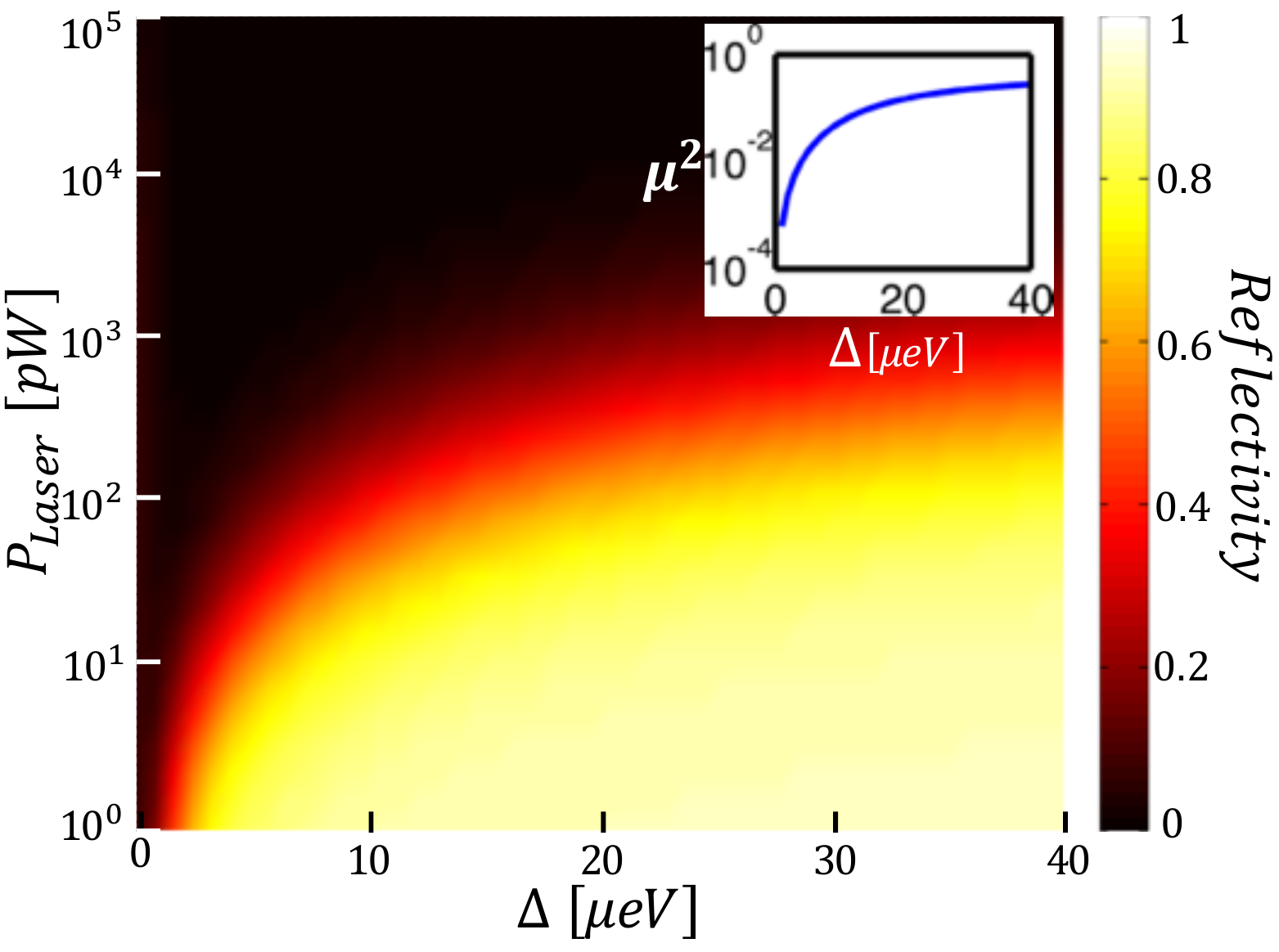}
	\caption{\label{fig:NL_vs_P_fx_detuning}Colormap of the reflectivity of the subradiant state $\ket{-''}$ as a function of the incident laser power and of the QDs detuning $\Delta_{12}=\frac{\omega_{1}-\omega_{2}}{2}$. Dipole-dipole interaction is kept the same with $\Omega_{12}=31~\si{\micro\electronvolt}$. The laser and cavity frequency are set to $\omega_{L}=\omega_{c}=\frac{\omega_{1}+\omega_{2}}{2}-\sqrt{\Delta_{12}	^2+\Omega_{12}	^{2}}$, in resonance with $\ket{-''}$. The subradiant state saturates at lower incident power for smaller detuning since its antisymmetric component is more pronounced. For very small detuning the state is completely antisymmetric and dark and doesn't couple to cavity. Inset: $\mu	^{2}$ the coefficient controlling the critical photon number for the subradiant state as a function of detuning. Parameters as in fig.6.}
\end{figure}

As mentioned in the introduction, the linewidth tunability is a very interesting property for interfacing bandwidth mismatched atomic systems. Quantum dots in micropillars possess bandwidths one or two order of magnitude larger than other common systems such as cold atoms, cold ions or superconducting circuits~\cite{shomroni2014,DeVoe1996,astafiev2010,ho2011}.
The strong tunability of the subradiant state's linewidth $\Gamma_{\ket{-''}}$, associated to the strong tunability of the coupling with the cavity mode $g_{\ket{-''}}$, allows to span the whole range of bandwidths smaller or equal than the one of the single quantum dot in a microcavity. This is performed by playing with the detuning between the two QDs, $\Delta_{12}$. Indeed Eq.~\ref{eq:Gamma-''} indicates that the bandwidth directly depends on $\mu^{2}$, which has been plotted in the inset of Fig.~\ref{fig:NL_vs_P_fx_detuning} as a function of $\Delta_{12}$. Importantly, only small changes in the detuning $\Delta_{12}$ are needed to obtain variations of several orders of magnitude of the bandwidth. Indeed, for detunings in the range $\Delta_{12}\in\left[3;\,50\right] \, \si{\micro\electronvolt}$, one obtains the bandwidths $\Gamma_{\ket{-''}}\in\left[0.038;\,3.8\right] \, \si{\micro\electronvolt}$: two orders of magnitude are available for a detuning of some tens of $\si{\micro\electronvolt}$.

\begin{figure}
	\includegraphics[width=.9\columnwidth]{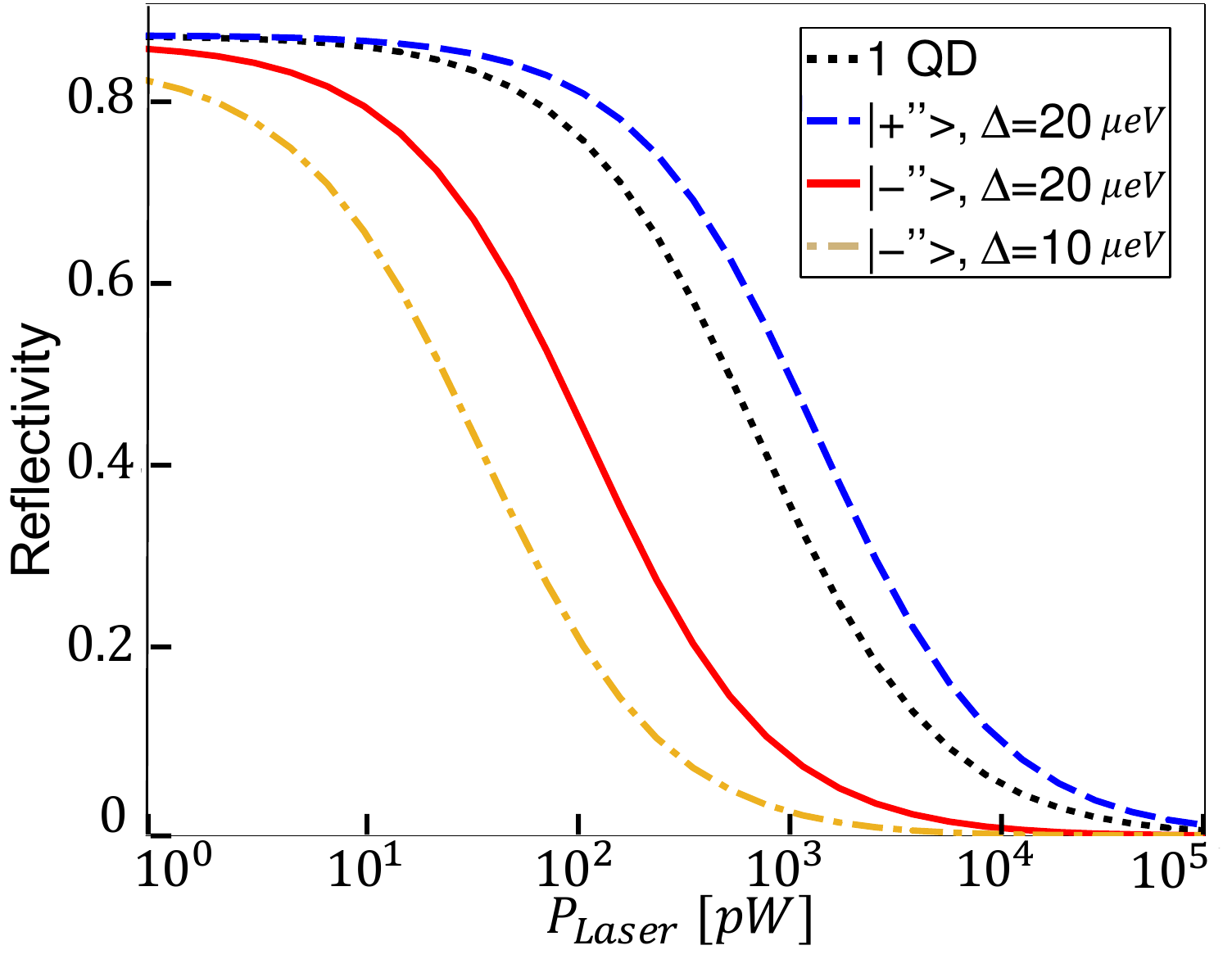}
	\caption{\label{fig:NL_vs_P}Reflectivity vs incident laser power. In dotted black for one quantum dot, in dashed blue for the superradiant state $\ket{+''}$ of two dipole-dipole coupled, detuned quantum dots with $\Delta_{12}=20\, \si{\micro\electronvolt}$, in solid red the subradiant state $\ket{-''}$ with the same detuning, and in dashed-dotted yellow the same state $\ket{-''}$ but for a smaller detuning of $\Delta_{12}=10\, \si{\micro\electronvolt}$. The parameters are $\left\{g,\kappa,\gamma\right\}=\left\{ 20, 200, 0.6 \right\}\,\si{\micro\electronvolt}$. The pump laser and the cavity mode are each time tuned to the frequency of the state in consideration.}
\end{figure}

The other important parameter that can be tuned is the nonlinear behavior of the superradiant and subradiant states $\ket{+''}$ and $\ket{-''}$. This is explored in more detail by calculating the reflectivity for a large range of incident laser powers. 
We plot the reflectivity of each state as a function of the incident laser power and compare it to the reference case of a single QD in a cavity. The laser frequency is now tuned to the energy of the state we want to study. We also tune the cavity mode to this frequency so that at very high power we obtain the same final reflectivity.
The results are shown in figure~\ref{fig:NL_vs_P} for QDs separated by $d=10\,\text{nm}$.
The black dotted line shows the response of the reference system with only one quantum dot. The reflectivity goes from close to one when the QD is not saturated to zero when the QD is completely bleached and the system behaves as an empty cavity and all the pump is transmitted. The superradiant state $\ket{+''}$, shown in dashed blue, is more robust to saturation and the reflectivity starts decreasing only at twice the incident power. On the contrary the subradiant state $\ket{-''}$ saturates easily and the nonlinear behavior is observed for an incident power almost an order of magnitude smaller for $\Delta_{12}=20\, \si{\micro\electronvolt}$. If now $\Delta_{12}=10\, \si{\micro\electronvolt}$ the non-linearity threshold of $\ket{-''}$ is reduced by a factor 3. This can be seen when comparing the solid red curve and the dotted dashed yellow curve representing these two cases.
Indeed, as the linewidth of the states, the saturation threshold depends directly on the detuning between the two quantum dots. This can be calculated with the critical photon number of Eq.~\ref{eq:critical_photon''}. As before the tunability comes from the modification of the coupling of the state to the cavity mode, but now it also depends on the modification of the free-space spontaneous emission $\gamma_{\ket{-''}}$ of Eq.~\ref{eq:spont''}. In the end, we get the same dependence on $\mu^{2}$ for the critical photon number than with the linewidth: $n_{c_{\ket{-''}}} = 2 \mu^{2} n_{c_{0}}$. 
To illustrate this, we plot in figure~\ref{fig:NL_vs_P_fx_detuning} the same nonlinear curves of figure~\ref{fig:NL_vs_P} but now as a function of $\Delta_{12}$ in the x-axis and the laser pump power in the y-axis.
For a given detuning, the reflectivity goes from $1$ to $0$ for increasing pump power as before, but the threshold changes as a function of the chosen detuning. For detunings much larger than $\Omega_{12}$ the system behaves as independent quantum dots. Accordingly the saturation threshold is the one observed for a single QD. As $\Delta_{12}$ is decreased the threshold of nonlinearity decreases over orders of magnitude.
\section{Bandwidth tunable photon blockade}
\begin{figure}
	\includegraphics[width=\columnwidth]{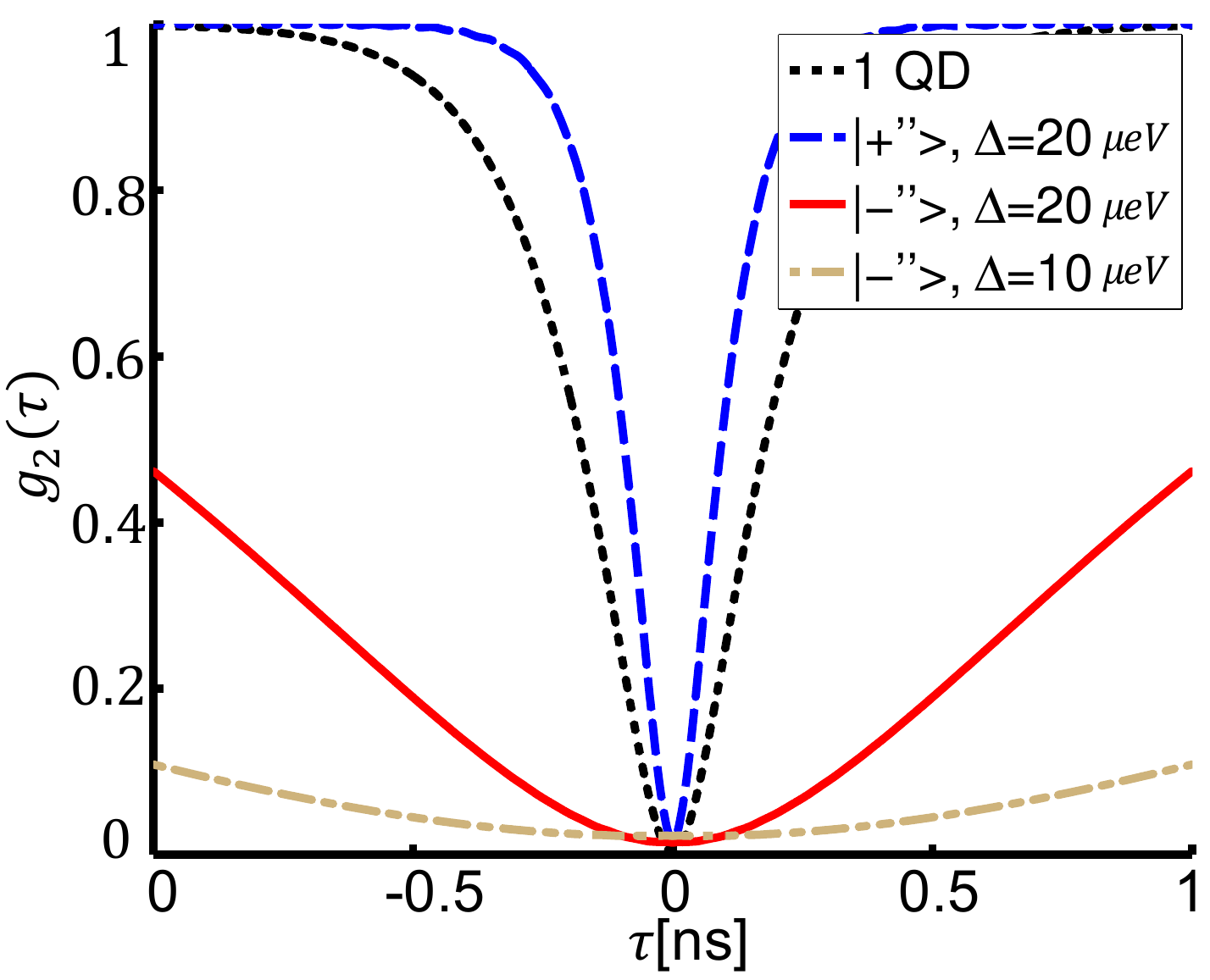}
	\caption{\label{fig:g2tau} Normalized second-order correlation function $g_{2}(\tau)$ as a function of delay between two detections. In dotted black for one quantum dot, in dashed blue for the superradiant state $\ket{+''}$ with $\Delta_{12}=20\,\si{\micro\electronvolt}$. The subradiant state $\ket{-''}$ has been shown at two different detunings $\Delta_{12}=20\,\si{\micro\electronvolt}$ in solid red and $\Delta_{12}=10\,\si{\micro\electronvolt}$ in dashed-dotted yellow. The cavity and laser frequency are matched to the respective state and $P_{laser}=10\,\text{pW}$. Parameters are the same as in fig.6.}
\end{figure}

The single photon non-linearity described above is central to the photon blockade phenomena, where a single photon conditions the transmission or reflection of a second one~\cite{shomroni2014, reiserer2014, volz2012,bennett2016,snijders2016, santis2017}. Such feature, which allows to build efficient photon-photon gates~\cite{shomroni2014,hacker2016}, can be evidenced through photon correlation measurements performed on a reflected laser beam on the device: a coherent light beam is converted into a subpoissonian light field showing anti-bunching over a time scale corresponding to the atomic transition optical strength~\cite{shomroni2014, tiecke2014, volz2012,bennett2016,snijders2016, santis2017}.  Indeed the reflected light results from the coherent superposition of the reflected laser and of the emitted light by the TLS through the cavity. Since we consider here a symmetric cavity, in the absence of the TLS and when $\omega_{L}=\omega_{c}$, no light from the laser is reflected: $\text{R}_{\text{min}}=0$. In the presence of the TLS, the light reflected is entirely due to the TLS emission and is thus antibunched. 
The poissonian statistics of the laser is converted in sub-poissonian, so the second-order correlation function of the reflected light at zero delay tends to zero: $g_{2}(0)=0$~\cite{Carmichael1976}.
The width of the correlation dip depends on the rate of re-excitation of the TLS through the cavity after having emitted one photon, which corresponds to the Purcell-enhanced linewidth $\Gamma$ of Eq.~\ref{eq:Gamma_purcell} and scales as $1/\Gamma$ in the low excitation limit~\cite{Walls2008}.
It is thus interesting to compare the response in correlation of the different coupled states of our system which behave as TLS with different linewidths $\Gamma$. As seen in Eq.~\ref{eq:Gamma+''} and Eq.~\ref{eq:Gamma-''}, the superradiant state $\ket{+''}$ has a cavity-enhanced emission rate $\Gamma_{\ket{+''}}$ close to twice as large as a single QD and the subradiant state $\ket{-''}$ has a cavity-modified emission rate $\Gamma_{\ket{-''}}$ ranging from 0 to $\Gamma_{0}$ depending strongly on the detuning $\Delta_{12}$. 
The normalized intensity correlation function~\cite{Walls2008}
\begin{equation}
g_{2}(\tau)=\frac{\langle a_{\text{out}}^{\dagger} (0) a_{\text{out}}^{\dagger}(\tau) a_{\text{out}}(\tau) a_{\text{out}}(0)\rangle}
{\langle a_{\text{out}}^{\dagger}a_{\text{out}}\rangle^{2}} \, ,
\end{equation}
with $a_{\text{out}}$ the reflected field operator, has been plotted for these two states and for the single QD case in figure~\ref{fig:g2tau}.
The cavity and laser frequencies are again matched to the state in consideration. All the curves show dips at zero delay but with different widths. 
As expected, the superradiant state has a thinner dip in correlation than the case of the single QD. On the contrary, the subradiant state has a larger dip than the single QD case. 
Importantly, the width of the dip can be adjusted by controlling the QD detuning $\Delta_{12}$. For example, when reducing the detuning from $\Delta_{12}=20\,\si{\micro\electronvolt}$ to $\Delta_{12}=10\,\si{\micro\electronvolt}$, it can be seen that the correlation dip of the $\ket{-''}$ state gets much wider and reaches a FWHM surpassing the nanosecond.
Such long-lived and detuning tunable antibunching shows that  the memory time of the system can be increased reaching values largely surpassing the nanosecond.

\ilan{\section{Concluding remarks}}
In summary, we have presented a theoretical study of two dipole-dipole coupled quantum dots interacting weakly with a laser-pumped cavity. We have found that detuning between the QDs allows to probe not only the superradiant state but also the subradiant state which is no longer dark. In the Purcell regime, only direct coupling allows to maintain collective behavior for detunings large enough so as to distinguish the different collective states. Having both detuning and direct coupling allows to independently take advantage of the superradiant and subradiant behaviors. In particular the non-linear behavior of the subradiant state can vary over order of magnitudes for very small changes in detuning. This was confirmed by calculations of the delayed second-order correlation function which showed that antibunching can be extended to very long delays by controlling the detuning between the QDs. The corresponding bandwidth tunability of the subradiant state can be useful to interface different quantum systems in view of building hybrid quantum networks where matching the spectral shape of an incoming pulse with the linewidth of the state is necessary to reach the maximum routing efficiency~\cite{Rosenblum2011}. 

As an example, QDs in micropillars, which are the most efficient single photon sources thus far~\cite{ding2016,somaschi2016}, present an optical bandwidth that is typically one or two orders of  magnitude larger than  cold atoms memories~\cite{shomroni2014}.  One  could envision to use the subradiant state of a coupled QD system to generate single photons matched to the smaller bandwidth of cold atoms or cold ions~\cite{DeVoe1996}. Another attractive possibility would be to use such coupled QDs system to store on a longer time scale short single photons emitted by bright QD based single photon sources~\cite{senellart2017}. To do this, one should excite the subradiant state for the detuned QDs case, choosing the appropriate detuning to match the incoming photon, and then, after the absorption, adiabatically bring them back to resonance. The stored photon could then be re-emitted when desired by detuning the QDs back again.

\begin{acknowledgments}
We thank N. Schilder for his precious help in the beginning of this study.
This work was supported by SAFRAN-IOGS chair on Ultimate Photonics, by the  ERC Starting Grant No. 277885 QD-CQED, the French Agence Nationale pour la Recherche grant SPIQE: ANR-14-CE32-0012; grant QDQN:ANR-17-ERC2-0014-01 and a public grant overseen by the French National Research Agency (ANR) as part of the "Investissements d'Avenir" program (Labex NanoSaclay, reference: ANR-10-LABX-0035), and the iXcore foundation. J.-J.G. acknowledges the support of Institut Universitaire de France.
\end{acknowledgments}

\appendix
\section{Effective Hamiltonian}\label{app:effective_hamiltonian}
The eigenstates of the system are obtained by studying the fluorescence emission spectra. 
Evolution of expectation values of operators defining the cavity field and the QDs dipole is calculated using the master Eq.~\ref{eq:equation-maitresse} and the property for a given operator $\hat{o}$ of the system $\ev{\dot{\hat{o}}}=\Tr(\hat{o} \dot{\rho})$. The coherent pump beam is replaced by incoherent pumping of the QDs or the cavity described by a term $P_{c}\mathcal{L}(\hat{a}^{\dagger})$ and  $P_{i}\mathcal{L}(\sigma^{+}_{i})$ for the the cavity or the i-th QD. $P_{c}$ and $P_{i}$ are the corresponding pump powers. We obtain:
\begin{align}\label{eq:ev_evol}
i\partialderivative{}{t}
\begin{pmatrix}
\ev{\sigma_{1}}\\
\ev{\sigma_{2}} \\
\ev{a} \\
\end{pmatrix}=
\begin{pmatrix}
\tilde{\omega}_{1} & \Omega_{12}-i\frac{\gamma_{12}}{2} & -ig\\
\Omega_{12}-i\frac{\gamma_{12}}{2} & \tilde{\omega}_{2} & -ig\\
ig & ig & \tilde{\omega}_{c}\\
\end{pmatrix}
\begin{pmatrix}
\ev{\sigma_{1}}\\
\ev{\sigma_{2}} \\
\ev{a} \\
\end{pmatrix}
\end{align}
where we have introduced the complex frequencies $\tilde{\omega}_{j}=\omega_{j}-i\frac{\gamma}{2}-iP_{j}$ and $\tilde{\omega}_{c}=\omega_{c}-i\frac{\kappa}{2}+iP_{c}$~\cite{Laussy2008}.
These equations are valid only at low excitation pump power (linear regime) with the approximation $\ev{\sigma_{z}}\simeq-1$ for each QD.
They are exact in the spontaneous emission case with at most one excitation in the system, restricted to the basis $\ket{0,g,g}$, $\ket{0,g,e}$, $\ket{0,e,g}$ and $\ket{1,g,g}$ where the first value in the ket corresponds to the number of photons in the cavity mode~\cite{Carmichael1989}.
The spectral function of the cavity is given by~\cite{Laucht2010}:
\begin{align}
	S(\omega)\propto\lim\limits_{t\rightarrow\infty}\text{Re}\int_{0}^{\infty}d\tau e^{-i\omega\tau}\ev{a^{\dagger}(t+\tau)a(t)}\, .
\end{align}
According to the quantum regression theorem, the evolution of the first order correlation function $\ev{a^{\dagger}(t+\tau)a(t)}$ obeys the same equations as $\ev{a}$. The eigenstates are then obtained by diagonalizing the matrix of Eq.~\ref{eq:ev_evol}. Fig.~\ref{fig:AB_fx_delta} and Fig.~\ref{fig:MuNu_fx_delta} present the components of the excitonic eigenvectors without or with dipole-dipole interaction.

\section{Cavity coupling rate}\label{app:cavity_coupling}
Interaction with the cavity is described by the term $ i\sqrt{2}g\left(a^{\dagger}\frac{(\sigma_{1}+\sigma_{2})}{\sqrt{2}}-a\frac{(\sigma_{1}^{+}+\sigma_{2}^{+})}{\sqrt{2}}\right)$ of the Hamiltonian of eq.~\ref{eq:hamiltonian}.
By rewriting the symmetric operator $\frac{\sigma_{1}+\sigma_{2}}{\sqrt{2}}$ as a function of the new coupled states we get:
\begin{align}\label{eq:sym_op}
\frac{\sigma_{1}+\sigma_{2}}{\sqrt{2}}&=
\nu\Big(\ketbra{+''}{ee}+\ketbra{gg}{+''}\Big)\nonumber\\
&+\mu\Big(\ketbra{gg}{-''}+\ketbra{-''}{ee}\Big)
\end{align}
where $\ketbra{i}{j}$ is the operator that transforms the state $j$ of the collective basis in the state $i$. 
We directly deduce the coupling strength of $\ket{-''}$ and $\ket{+''}$ to the cavity mode: 
\begin{align}
g_{\ket{-''}}&= \mu \sqrt{2}g\\
g_{\ket{+''}}&= \nu \sqrt{2}g \, .
\end{align}

\section{Spontaneous emission rate}
The spontaneous emission rate of the new states is calculated in the same way. The Lindbladian describing spontaneous emission of the QDs is:
\begin{align} \label{eq:lind_sym}
	(\gamma+\gamma_{12})\mathcal{L}\left(\frac{\sigma_{1}+\sigma_{2}}{\sqrt{2}}\right) \, .
\end{align}
We neglect here the antisymmetric terms by considering $\gamma-\gamma_{12}\simeq0$.
When injecting the expression of the symmetric operator~\ref{eq:sym_op} in Eq.~\ref{eq:lind_sym} we obtain:
\begin{align*}
&(\gamma+\gamma_{12})
\Big(\nu^{2}\mathcal{L}\big(\ketbra{+''}{ee}+\ketbra{gg}{+''}\big)\\
&+\mu^{2}\mathcal{L}\big(\ketbra{-''}{ee}+\ketbra{gg}{-''}\big)\\
&+\mu\nu \big(\ketbra{gg}{-''} + \ketbra{-''}{ee}\big)\rho\big(\ketbra{ee}{+''} + \ketbra{+''}{gg}\big)\\
&+\mu\nu \big(\ketbra{gg}{+''} + \ketbra{+''}{ee}\big)\rho\big(\ketbra{ee}{-''} + \ketbra{-''}{gg}\big) \\
&+\frac{1}{2}\mu\nu\big( \ketbra{+''}{-''} - \ketbra{-''}{+''} \big)\rho \\
&+\frac{1}{2}\mu\nu\rho \big( \ketbra{+''}{-''} - \ketbra{-''}{+''} \big)
\Big)
\, .
\end{align*}
The first term is the decay through the symmetric branch $\ket{ee}\rightarrow \ket{+''} \rightarrow \ket{gg}$ with a rate $\nu	^{2}(\gamma+\gamma_{12})$, the second term is the decay through the antisymmetric branch $\ket{ee}\rightarrow \ket{-''} \rightarrow \ket{gg}$
with a (negligible) rate $\mu^{2}(\gamma+\gamma_{12})$ and the last term are cross terms that couple incoherently the entangled states with a rate $\mu\nu(\gamma+\gamma_{12})$.


\begin{thebibliography}{59}%
	\makeatletter
	\providecommand \@ifxundefined [1]{%
		\@ifx{#1\undefined}
	}%
	\providecommand \@ifnum [1]{%
		\ifnum #1\expandafter \@firstoftwo
		\else \expandafter \@secondoftwo
		\fi
	}%
	\providecommand \@ifx [1]{%
		\ifx #1\expandafter \@firstoftwo
		\else \expandafter \@secondoftwo
		\fi
	}%
	\providecommand \natexlab [1]{#1}%
	\providecommand \enquote  [1]{``#1''}%
	\providecommand \bibnamefont  [1]{#1}%
	\providecommand \bibfnamefont [1]{#1}%
	\providecommand \citenamefont [1]{#1}%
	\providecommand \href@noop [0]{\@secondoftwo}%
	\providecommand \href [0]{\begingroup \@sanitize@url \@href}%
	\providecommand \@href[1]{\@@startlink{#1}\@@href}%
	\providecommand \@@href[1]{\endgroup#1\@@endlink}%
	\providecommand \@sanitize@url [0]{\catcode `\\12\catcode `\$12\catcode
		`\&12\catcode `\#12\catcode `\^12\catcode `\_12\catcode `\%12\relax}%
	\providecommand \@@startlink[1]{}%
	\providecommand \@@endlink[0]{}%
	\providecommand \url  [0]{\begingroup\@sanitize@url \@url }%
	\providecommand \@url [1]{\endgroup\@href {#1}{\urlprefix }}%
	\providecommand \urlprefix  [0]{URL }%
	\providecommand \Eprint [0]{\href }%
	\providecommand \doibase [0]{http://dx.doi.org/}%
	\providecommand \selectlanguage [0]{\@gobble}%
	\providecommand \bibinfo  [0]{\@secondoftwo}%
	\providecommand \bibfield  [0]{\@secondoftwo}%
	\providecommand \translation [1]{[#1]}%
	\providecommand \BibitemOpen [0]{}%
	\providecommand \bibitemStop [0]{}%
	\providecommand \bibitemNoStop [0]{.\EOS\space}%
	\providecommand \EOS [0]{\spacefactor3000\relax}%
	\providecommand \BibitemShut  [1]{\csname bibitem#1\endcsname}%
	\let\auto@bib@innerbib\@empty
	\bibitem [{\citenamefont {Knill}\ \emph {et~al.}(2001)\citenamefont {Knill},
		\citenamefont {Laflamme},\ and\ \citenamefont {Milburn}}]{knill2001}%
	\BibitemOpen
	\bibfield  {author} {\bibinfo {author} {\bibfnamefont {E.}~\bibnamefont
			{Knill}}, \bibinfo {author} {\bibfnamefont {R.}~\bibnamefont {Laflamme}}, \
		and\ \bibinfo {author} {\bibfnamefont {G.~J.}\ \bibnamefont {Milburn}},\
	}\href {\doibase 10.1038/35051009} {\bibfield  {journal} {\bibinfo  {journal}
			{Nature}\ }\textbf {\bibinfo {volume} {409}},\ \bibinfo {pages} {46}
		(\bibinfo {year} {2001})}\BibitemShut {NoStop}%
	\bibitem [{\citenamefont {Briegel}\ \emph {et~al.}(1998)\citenamefont
		{Briegel}, \citenamefont {Dür}, \citenamefont {Cirac},\ and\ \citenamefont
		{Zoller}}]{briegel1998}%
	\BibitemOpen
	\bibfield  {author} {\bibinfo {author} {\bibfnamefont {H.-J.}\ \bibnamefont
			{Briegel}}, \bibinfo {author} {\bibfnamefont {W.}~\bibnamefont {Dür}},
		\bibinfo {author} {\bibfnamefont {J.~I.}\ \bibnamefont {Cirac}}, \ and\
		\bibinfo {author} {\bibfnamefont {P.}~\bibnamefont {Zoller}},\ }\href
	{\doibase 10.1103/PhysRevLett.81.5932} {\bibfield  {journal} {\bibinfo
			{journal} {Physical Review Letters}\ }\textbf {\bibinfo {volume} {81}},\
		\bibinfo {pages} {5932} (\bibinfo {year} {1998})}\BibitemShut {NoStop}%
	\bibitem [{\citenamefont {Kimble}(2008)}]{kimble2008}%
	\BibitemOpen
	\bibfield  {author} {\bibinfo {author} {\bibfnamefont {H.~J.}\ \bibnamefont
			{Kimble}},\ }\href {\doibase 10.1038/nature07127} {\bibfield  {journal}
		{\bibinfo  {journal} {Nature}\ }\textbf {\bibinfo {volume} {453}},\ \bibinfo
		{pages} {1023} (\bibinfo {year} {2008})}\BibitemShut {NoStop}%
	\bibitem [{\citenamefont {Shomroni}\ \emph {et~al.}(2014)\citenamefont
		{Shomroni}, \citenamefont {Rosenblum}, \citenamefont {Lovsky}, \citenamefont
		{Bechler}, \citenamefont {Guendelman},\ and\ \citenamefont
		{Dayan}}]{shomroni2014}%
	\BibitemOpen
	\bibfield  {author} {\bibinfo {author} {\bibfnamefont {I.}~\bibnamefont
			{Shomroni}}, \bibinfo {author} {\bibfnamefont {S.}~\bibnamefont {Rosenblum}},
		\bibinfo {author} {\bibfnamefont {Y.}~\bibnamefont {Lovsky}}, \bibinfo
		{author} {\bibfnamefont {O.}~\bibnamefont {Bechler}}, \bibinfo {author}
		{\bibfnamefont {G.}~\bibnamefont {Guendelman}}, \ and\ \bibinfo {author}
		{\bibfnamefont {B.}~\bibnamefont {Dayan}},\ }\href {\doibase
		10.1126/science.1254699} {\bibfield  {journal} {\bibinfo  {journal}
			{Science}\ }\textbf {\bibinfo {volume} {345}},\ \bibinfo {pages} {903}
		(\bibinfo {year} {2014})}\BibitemShut {NoStop}%
	\bibitem [{\citenamefont {Reiserer}\ \emph {et~al.}(2014)\citenamefont
		{Reiserer}, \citenamefont {Kalb}, \citenamefont {Rempe},\ and\ \citenamefont
		{Ritter}}]{reiserer2014}%
	\BibitemOpen
	\bibfield  {author} {\bibinfo {author} {\bibfnamefont {A.}~\bibnamefont
			{Reiserer}}, \bibinfo {author} {\bibfnamefont {N.}~\bibnamefont {Kalb}},
		\bibinfo {author} {\bibfnamefont {G.}~\bibnamefont {Rempe}}, \ and\ \bibinfo
		{author} {\bibfnamefont {S.}~\bibnamefont {Ritter}},\ }\href {\doibase
		10.1038/nature13177} {\bibfield  {journal} {\bibinfo  {journal} {Nature}\
		}\textbf {\bibinfo {volume} {508}},\ \bibinfo {pages} {237} (\bibinfo {year}
		{2014})}\BibitemShut {NoStop}%
	\bibitem [{\citenamefont {Hacker}\ \emph {et~al.}(2016)\citenamefont {Hacker},
		\citenamefont {Welte}, \citenamefont {Rempe},\ and\ \citenamefont
		{Ritter}}]{hacker2016}%
	\BibitemOpen
	\bibfield  {author} {\bibinfo {author} {\bibfnamefont {B.}~\bibnamefont
			{Hacker}}, \bibinfo {author} {\bibfnamefont {S.}~\bibnamefont {Welte}},
		\bibinfo {author} {\bibfnamefont {G.}~\bibnamefont {Rempe}}, \ and\ \bibinfo
		{author} {\bibfnamefont {S.}~\bibnamefont {Ritter}},\ }\href {\doibase
		10.1038/nature18592} {\bibfield  {journal} {\bibinfo  {journal} {Nature}\
		}\textbf {\bibinfo {volume} {536}},\ \bibinfo {pages} {193} (\bibinfo {year}
		{2016})}\BibitemShut {NoStop}%
	\bibitem [{\citenamefont {Volz}\ \emph {et~al.}(2012)\citenamefont {Volz},
		\citenamefont {Reinhard}, \citenamefont {Winger}, \citenamefont {Badolato},
		\citenamefont {Hennessy}, \citenamefont {Hu},\ and\ \citenamefont
		{Imamoğlu}}]{volz2012}%
	\BibitemOpen
	\bibfield  {author} {\bibinfo {author} {\bibfnamefont {T.}~\bibnamefont
			{Volz}}, \bibinfo {author} {\bibfnamefont {A.}~\bibnamefont {Reinhard}},
		\bibinfo {author} {\bibfnamefont {M.}~\bibnamefont {Winger}}, \bibinfo
		{author} {\bibfnamefont {A.}~\bibnamefont {Badolato}}, \bibinfo {author}
		{\bibfnamefont {K.~J.}\ \bibnamefont {Hennessy}}, \bibinfo {author}
		{\bibfnamefont {E.~L.}\ \bibnamefont {Hu}}, \ and\ \bibinfo {author}
		{\bibfnamefont {A.}~\bibnamefont {Imamoğlu}},\ }\href {\doibase
		10.1038/nphoton.2012.181} {\bibfield  {journal} {\bibinfo  {journal} {Nature
				Photonics}\ }\textbf {\bibinfo {volume} {6}},\ \bibinfo {pages} {605}
		(\bibinfo {year} {2012})}\BibitemShut {NoStop}%
	\bibitem [{\citenamefont {Bonato}\ \emph {et~al.}(2010)\citenamefont {Bonato},
		\citenamefont {Haupt}, \citenamefont {Oemrawsingh}, \citenamefont {Gudat},
		\citenamefont {Ding}, \citenamefont {vanExter},\ and\ \citenamefont
		{Bouwmeester}}]{bonato2010}%
	\BibitemOpen
	\bibfield  {author} {\bibinfo {author} {\bibfnamefont {C.}~\bibnamefont
			{Bonato}}, \bibinfo {author} {\bibfnamefont {F.}~\bibnamefont {Haupt}},
		\bibinfo {author} {\bibfnamefont {S.~S.~R.}\ \bibnamefont {Oemrawsingh}},
		\bibinfo {author} {\bibfnamefont {J.}~\bibnamefont {Gudat}}, \bibinfo
		{author} {\bibfnamefont {D.}~\bibnamefont {Ding}}, \bibinfo {author}
		{\bibfnamefont {M.~P.}\ \bibnamefont {vanExter}}, \ and\ \bibinfo {author}
		{\bibfnamefont {D.}~\bibnamefont {Bouwmeester}},\ }\href {\doibase
		10.1103/PhysRevLett.104.160503} {\bibfield  {journal} {\bibinfo  {journal}
			{Physical Review Letters}\ }\textbf {\bibinfo {volume} {104}},\ \bibinfo
		{pages} {160503} (\bibinfo {year} {2010})}\BibitemShut {NoStop}%
	\bibitem [{\citenamefont {Julsgaard}\ \emph {et~al.}(2004)\citenamefont
		{Julsgaard}, \citenamefont {Sherson}, \citenamefont {Cirac}, \citenamefont
		{Fiurášek},\ and\ \citenamefont {Polzik}}]{julsgaard2004}%
	\BibitemOpen
	\bibfield  {author} {\bibinfo {author} {\bibfnamefont {B.}~\bibnamefont
			{Julsgaard}}, \bibinfo {author} {\bibfnamefont {J.}~\bibnamefont {Sherson}},
		\bibinfo {author} {\bibfnamefont {J.~I.}\ \bibnamefont {Cirac}}, \bibinfo
		{author} {\bibfnamefont {J.}~\bibnamefont {Fiurášek}}, \ and\ \bibinfo
		{author} {\bibfnamefont {E.~S.}\ \bibnamefont {Polzik}},\ }\href {\doibase
		10.1038/nature03064} {\bibfield  {journal} {\bibinfo  {journal} {Nature}\
		}\textbf {\bibinfo {volume} {432}},\ \bibinfo {pages} {482} (\bibinfo {year}
		{2004})}\BibitemShut {NoStop}%
	\bibitem [{\citenamefont {Chanelière}\ \emph {et~al.}(2005)\citenamefont
		{Chanelière}, \citenamefont {Matsukevich}, \citenamefont {Jenkins},
		\citenamefont {Lan}, \citenamefont {Kennedy},\ and\ \citenamefont
		{Kuzmich}}]{chaneliere2005}%
	\BibitemOpen
	\bibfield  {author} {\bibinfo {author} {\bibfnamefont {T.}~\bibnamefont
			{Chanelière}}, \bibinfo {author} {\bibfnamefont {D.~N.}\ \bibnamefont
			{Matsukevich}}, \bibinfo {author} {\bibfnamefont {S.~D.}\ \bibnamefont
			{Jenkins}}, \bibinfo {author} {\bibfnamefont {S.-Y.}\ \bibnamefont {Lan}},
		\bibinfo {author} {\bibfnamefont {T.~a.~B.}\ \bibnamefont {Kennedy}}, \ and\
		\bibinfo {author} {\bibfnamefont {A.}~\bibnamefont {Kuzmich}},\ }\href
	{\doibase 10.1038/nature04315} {\bibfield  {journal} {\bibinfo  {journal}
			{Nature}\ }\textbf {\bibinfo {volume} {438}},\ \bibinfo {pages} {833}
		(\bibinfo {year} {2005})}\BibitemShut {NoStop}%
	\bibitem [{\citenamefont {Sayrin}\ \emph {et~al.}(2015)\citenamefont {Sayrin},
		\citenamefont {Clausen}, \citenamefont {Albrecht}, \citenamefont
		{Schneeweiss},\ and\ \citenamefont {Rauschenbeutel}}]{sayrin2015}%
	\BibitemOpen
	\bibfield  {author} {\bibinfo {author} {\bibfnamefont {C.}~\bibnamefont
			{Sayrin}}, \bibinfo {author} {\bibfnamefont {C.}~\bibnamefont {Clausen}},
		\bibinfo {author} {\bibfnamefont {B.}~\bibnamefont {Albrecht}}, \bibinfo
		{author} {\bibfnamefont {P.}~\bibnamefont {Schneeweiss}}, \ and\ \bibinfo
		{author} {\bibfnamefont {A.}~\bibnamefont {Rauschenbeutel}},\ }\href
	{\doibase 10.1364/OPTICA.2.000353} {\bibfield  {journal} {\bibinfo  {journal}
			{Optica}\ }\textbf {\bibinfo {volume} {2}},\ \bibinfo {pages} {353} (\bibinfo
		{year} {2015})}\BibitemShut {NoStop}%
	\bibitem [{\citenamefont {Specht}\ \emph {et~al.}(2011)\citenamefont {Specht},
		\citenamefont {Nölleke}, \citenamefont {Reiserer}, \citenamefont {Uphoff},
		\citenamefont {Figueroa}, \citenamefont {Ritter},\ and\ \citenamefont
		{Rempe}}]{specht2011}%
	\BibitemOpen
	\bibfield  {author} {\bibinfo {author} {\bibfnamefont {H.~P.}\ \bibnamefont
			{Specht}}, \bibinfo {author} {\bibfnamefont {C.}~\bibnamefont {Nölleke}},
		\bibinfo {author} {\bibfnamefont {A.}~\bibnamefont {Reiserer}}, \bibinfo
		{author} {\bibfnamefont {M.}~\bibnamefont {Uphoff}}, \bibinfo {author}
		{\bibfnamefont {E.}~\bibnamefont {Figueroa}}, \bibinfo {author}
		{\bibfnamefont {S.}~\bibnamefont {Ritter}}, \ and\ \bibinfo {author}
		{\bibfnamefont {G.}~\bibnamefont {Rempe}},\ }\href {\doibase
		10.1038/nature09997} {\bibfield  {journal} {\bibinfo  {journal} {Nature}\
		}\textbf {\bibinfo {volume} {473}},\ \bibinfo {pages} {190} (\bibinfo {year}
		{2011})}\BibitemShut {NoStop}%
	\bibitem [{\citenamefont {Choi}\ \emph {et~al.}(2008)\citenamefont {Choi},
		\citenamefont {Deng}, \citenamefont {Laurat},\ and\ \citenamefont
		{Kimble}}]{choi2008}%
	\BibitemOpen
	\bibfield  {author} {\bibinfo {author} {\bibfnamefont {K.~S.}\ \bibnamefont
			{Choi}}, \bibinfo {author} {\bibfnamefont {H.}~\bibnamefont {Deng}}, \bibinfo
		{author} {\bibfnamefont {J.}~\bibnamefont {Laurat}}, \ and\ \bibinfo {author}
		{\bibfnamefont {H.~J.}\ \bibnamefont {Kimble}},\ }\href {\doibase
		10.1038/nature06670} {\bibfield  {journal} {\bibinfo  {journal} {Nature}\
		}\textbf {\bibinfo {volume} {452}},\ \bibinfo {pages} {67} (\bibinfo {year}
		{2008})}\BibitemShut {NoStop}%
	\bibitem [{\citenamefont {Körber}\ \emph {et~al.}(2018)\citenamefont
		{Körber}, \citenamefont {Morin}, \citenamefont {Langenfeld}, \citenamefont
		{Neuzner}, \citenamefont {Ritter},\ and\ \citenamefont {Rempe}}]{korber2018}%
	\BibitemOpen
	\bibfield  {author} {\bibinfo {author} {\bibfnamefont {M.}~\bibnamefont
			{Körber}}, \bibinfo {author} {\bibfnamefont {O.}~\bibnamefont {Morin}},
		\bibinfo {author} {\bibfnamefont {S.}~\bibnamefont {Langenfeld}}, \bibinfo
		{author} {\bibfnamefont {A.}~\bibnamefont {Neuzner}}, \bibinfo {author}
		{\bibfnamefont {S.}~\bibnamefont {Ritter}}, \ and\ \bibinfo {author}
		{\bibfnamefont {G.}~\bibnamefont {Rempe}},\ }\href {\doibase
		10.1038/s41566-017-0050-y} {\bibfield  {journal} {\bibinfo  {journal} {Nature
				Photonics}\ }\textbf {\bibinfo {volume} {12}},\ \bibinfo {pages} {18}
		(\bibinfo {year} {2018})}\BibitemShut {NoStop}%
	\bibitem [{\citenamefont {Dayan}\ \emph {et~al.}(2008)\citenamefont {Dayan},
		\citenamefont {Parkins}, \citenamefont {Aoki}, \citenamefont {Ostby},
		\citenamefont {Vahala},\ and\ \citenamefont {Kimble}}]{dayan2008}%
	\BibitemOpen
	\bibfield  {author} {\bibinfo {author} {\bibfnamefont {B.}~\bibnamefont
			{Dayan}}, \bibinfo {author} {\bibfnamefont {A.~S.}\ \bibnamefont {Parkins}},
		\bibinfo {author} {\bibfnamefont {T.}~\bibnamefont {Aoki}}, \bibinfo {author}
		{\bibfnamefont {E.~P.}\ \bibnamefont {Ostby}}, \bibinfo {author}
		{\bibfnamefont {K.~J.}\ \bibnamefont {Vahala}}, \ and\ \bibinfo {author}
		{\bibfnamefont {H.~J.}\ \bibnamefont {Kimble}},\ }\href {\doibase
		10.1126/science.1152261} {\bibfield  {journal} {\bibinfo  {journal}
			{Science}\ }\textbf {\bibinfo {volume} {319}},\ \bibinfo {pages} {1062}
		(\bibinfo {year} {2008})}\BibitemShut {NoStop}%
	\bibitem [{\citenamefont {Loo}\ \emph {et~al.}(2012)\citenamefont {Loo},
		\citenamefont {Arnold}, \citenamefont {Gazzano}, \citenamefont {Lemaitre},
		\citenamefont {Sagnes}, \citenamefont {Krebs}, \citenamefont {Voisin},
		\citenamefont {Senellart},\ and\ \citenamefont {Lanco}}]{loo2012}%
	\BibitemOpen
	\bibfield  {author} {\bibinfo {author} {\bibfnamefont {V.}~\bibnamefont
			{Loo}}, \bibinfo {author} {\bibfnamefont {C.}~\bibnamefont {Arnold}},
		\bibinfo {author} {\bibfnamefont {O.}~\bibnamefont {Gazzano}}, \bibinfo
		{author} {\bibfnamefont {A.}~\bibnamefont {Lemaitre}}, \bibinfo {author}
		{\bibfnamefont {I.}~\bibnamefont {Sagnes}}, \bibinfo {author} {\bibfnamefont
			{O.}~\bibnamefont {Krebs}}, \bibinfo {author} {\bibfnamefont
			{P.}~\bibnamefont {Voisin}}, \bibinfo {author} {\bibfnamefont
			{P.}~\bibnamefont {Senellart}}, \ and\ \bibinfo {author} {\bibfnamefont
			{L.}~\bibnamefont {Lanco}},\ }\href {\doibase 10.1103/PhysRevLett.109.166806}
	{\bibfield  {journal} {\bibinfo  {journal} {Physical Review Letters}\
		}\textbf {\bibinfo {volume} {109}},\ \bibinfo {pages} {166806} (\bibinfo
		{year} {2012})}\BibitemShut {NoStop}%
	\bibitem [{\citenamefont {Bose}\ \emph {et~al.}(2012)\citenamefont {Bose},
		\citenamefont {Sridharan}, \citenamefont {Kim}, \citenamefont {Solomon},\
		and\ \citenamefont {Waks}}]{bose2012}%
	\BibitemOpen
	\bibfield  {author} {\bibinfo {author} {\bibfnamefont {R.}~\bibnamefont
			{Bose}}, \bibinfo {author} {\bibfnamefont {D.}~\bibnamefont {Sridharan}},
		\bibinfo {author} {\bibfnamefont {H.}~\bibnamefont {Kim}}, \bibinfo {author}
		{\bibfnamefont {G.~S.}\ \bibnamefont {Solomon}}, \ and\ \bibinfo {author}
		{\bibfnamefont {E.}~\bibnamefont {Waks}},\ }\href {\doibase
		10.1103/PhysRevLett.108.227402} {\bibfield  {journal} {\bibinfo  {journal}
			{Physical Review Letters}\ }\textbf {\bibinfo {volume} {108}},\ \bibinfo
		{pages} {227402} (\bibinfo {year} {2012})}\BibitemShut {NoStop}%
	\bibitem [{\citenamefont {Ding}\ \emph {et~al.}(2016)\citenamefont {Ding},
		\citenamefont {He}, \citenamefont {Duan}, \citenamefont {Gregersen},
		\citenamefont {Chen}, \citenamefont {Unsleber}, \citenamefont {Maier},
		\citenamefont {Schneider}, \citenamefont {Kamp}, \citenamefont {Hofling},
		\citenamefont {Lu},\ and\ \citenamefont {Pan}}]{ding2016}%
	\BibitemOpen
	\bibfield  {author} {\bibinfo {author} {\bibfnamefont {X.}~\bibnamefont
			{Ding}}, \bibinfo {author} {\bibfnamefont {Y.}~\bibnamefont {He}}, \bibinfo
		{author} {\bibfnamefont {Z.-C.}\ \bibnamefont {Duan}}, \bibinfo {author}
		{\bibfnamefont {N.}~\bibnamefont {Gregersen}}, \bibinfo {author}
		{\bibfnamefont {M.-C.}\ \bibnamefont {Chen}}, \bibinfo {author}
		{\bibfnamefont {S.}~\bibnamefont {Unsleber}}, \bibinfo {author}
		{\bibfnamefont {S.}~\bibnamefont {Maier}}, \bibinfo {author} {\bibfnamefont
			{C.}~\bibnamefont {Schneider}}, \bibinfo {author} {\bibfnamefont
			{M.}~\bibnamefont {Kamp}}, \bibinfo {author} {\bibfnamefont {S.}~\bibnamefont
			{Hofling}}, \bibinfo {author} {\bibfnamefont {C.-Y.}\ \bibnamefont {Lu}}, \
		and\ \bibinfo {author} {\bibfnamefont {J.-W.}\ \bibnamefont {Pan}},\ }\href
	{\doibase 10.1103/PhysRevLett.116.020401} {\bibfield  {journal} {\bibinfo
			{journal} {Physical Review Letters}\ }\textbf {\bibinfo {volume} {116}},\
		\bibinfo {pages} {020401} (\bibinfo {year} {2016})}\BibitemShut {NoStop}%
	\bibitem [{\citenamefont {Somaschi}\ \emph {et~al.}(2016)\citenamefont
		{Somaschi}, \citenamefont {Giesz}, \citenamefont {Santis}, \citenamefont
		{Loredo}, \citenamefont {Almeida}, \citenamefont {Hornecker}, \citenamefont
		{Portalupi}, \citenamefont {Grange}, \citenamefont {Antón}, \citenamefont
		{Demory}, \citenamefont {Gómez}, \citenamefont {Sagnes}, \citenamefont
		{Lanzillotti-Kimura}, \citenamefont {Lemaítre}, \citenamefont {Auffeves},
		\citenamefont {White}, \citenamefont {Lanco},\ and\ \citenamefont
		{Senellart}}]{somaschi2016}%
	\BibitemOpen
	\bibfield  {author} {\bibinfo {author} {\bibfnamefont {N.}~\bibnamefont
			{Somaschi}}, \bibinfo {author} {\bibfnamefont {V.}~\bibnamefont {Giesz}},
		\bibinfo {author} {\bibfnamefont {L.~D.}\ \bibnamefont {Santis}}, \bibinfo
		{author} {\bibfnamefont {J.~C.}\ \bibnamefont {Loredo}}, \bibinfo {author}
		{\bibfnamefont {M.~P.}\ \bibnamefont {Almeida}}, \bibinfo {author}
		{\bibfnamefont {G.}~\bibnamefont {Hornecker}}, \bibinfo {author}
		{\bibfnamefont {S.~L.}\ \bibnamefont {Portalupi}}, \bibinfo {author}
		{\bibfnamefont {T.}~\bibnamefont {Grange}}, \bibinfo {author} {\bibfnamefont
			{C.}~\bibnamefont {Antón}}, \bibinfo {author} {\bibfnamefont
			{J.}~\bibnamefont {Demory}}, \bibinfo {author} {\bibfnamefont
			{C.}~\bibnamefont {Gómez}}, \bibinfo {author} {\bibfnamefont
			{I.}~\bibnamefont {Sagnes}}, \bibinfo {author} {\bibfnamefont {N.~D.}\
			\bibnamefont {Lanzillotti-Kimura}}, \bibinfo {author} {\bibfnamefont
			{A.}~\bibnamefont {Lemaítre}}, \bibinfo {author} {\bibfnamefont
			{A.}~\bibnamefont {Auffeves}}, \bibinfo {author} {\bibfnamefont {A.~G.}\
			\bibnamefont {White}}, \bibinfo {author} {\bibfnamefont {L.}~\bibnamefont
			{Lanco}}, \ and\ \bibinfo {author} {\bibfnamefont {P.}~\bibnamefont
			{Senellart}},\ }\href {\doibase 10.1038/nphoton.2016.23} {\bibfield
		{journal} {\bibinfo  {journal} {Nature Photonics}\ }\textbf {\bibinfo
			{volume} {10}},\ \bibinfo {pages} {340} (\bibinfo {year} {2016})}\BibitemShut
	{NoStop}%
	\bibitem [{\citenamefont {Senellart}\ \emph {et~al.}(2017)\citenamefont
		{Senellart}, \citenamefont {Solomon},\ and\ \citenamefont
		{White}}]{senellart2017}%
	\BibitemOpen
	\bibfield  {author} {\bibinfo {author} {\bibfnamefont {P.}~\bibnamefont
			{Senellart}}, \bibinfo {author} {\bibfnamefont {G.}~\bibnamefont {Solomon}},
		\ and\ \bibinfo {author} {\bibfnamefont {A.}~\bibnamefont {White}},\ }\href
	{\doibase 10.1038/nnano.2017.218} {\bibfield  {journal} {\bibinfo  {journal}
			{Nature Nanotechnology}\ }\textbf {\bibinfo {volume} {12}},\ \bibinfo {pages}
		{1026} (\bibinfo {year} {2017})}\BibitemShut {NoStop}%
	\bibitem [{\citenamefont {Auffeves-Garnier}\ \emph {et~al.}(2007)\citenamefont
		{Auffeves-Garnier}, \citenamefont {Simon}, \citenamefont {Gerard},\ and\
		\citenamefont {Poizat}}]{auffeves2007}%
	\BibitemOpen
	\bibfield  {author} {\bibinfo {author} {\bibfnamefont {A.}~\bibnamefont
			{Auffeves-Garnier}}, \bibinfo {author} {\bibfnamefont {C.}~\bibnamefont
			{Simon}}, \bibinfo {author} {\bibfnamefont {J.-M.}\ \bibnamefont {Gerard}}, \
		and\ \bibinfo {author} {\bibfnamefont {J.-P.}\ \bibnamefont {Poizat}},\
	}\href {\doibase 10.1103/PhysRevA.75.053823} {\bibfield  {journal} {\bibinfo
			{journal} {Physical Review A}\ }\textbf {\bibinfo {volume} {75}},\ \bibinfo
		{pages} {053823} (\bibinfo {year} {2007})},\ \bibinfo {note}
	{wOS:000246890400168}\BibitemShut {NoStop}%
	\bibitem [{\citenamefont {Gardiner}\ and\ \citenamefont
		{Collett}(1985)}]{Gardiner1985}%
	\BibitemOpen
	\bibfield  {author} {\bibinfo {author} {\bibfnamefont {C.~W.}\ \bibnamefont
			{Gardiner}}\ and\ \bibinfo {author} {\bibfnamefont {M.~J.}\ \bibnamefont
			{Collett}},\ }\href {\doibase 10.1103/PhysRevA.31.3761} {\bibfield  {journal}
		{\bibinfo  {journal} {Phys. Rev. A}\ }\textbf {\bibinfo {volume} {31}},\
		\bibinfo {pages} {3761} (\bibinfo {year} {1985})}\BibitemShut {NoStop}%
	\bibitem [{\citenamefont {Tiecke}\ \emph {et~al.}(2014)\citenamefont {Tiecke},
		\citenamefont {Thompson}, \citenamefont {Leon}, \citenamefont {Liu},
		\citenamefont {Vuletić},\ and\ \citenamefont {Lukin}}]{tiecke2014}%
	\BibitemOpen
	\bibfield  {author} {\bibinfo {author} {\bibfnamefont {T.~G.}\ \bibnamefont
			{Tiecke}}, \bibinfo {author} {\bibfnamefont {J.~D.}\ \bibnamefont
			{Thompson}}, \bibinfo {author} {\bibfnamefont {N.~P.}\ \bibnamefont {Leon}},
		\bibinfo {author} {\bibfnamefont {L.~R.}\ \bibnamefont {Liu}}, \bibinfo
		{author} {\bibfnamefont {V.}~\bibnamefont {Vuletić}}, \ and\ \bibinfo
		{author} {\bibfnamefont {M.~D.}\ \bibnamefont {Lukin}},\ }\href {\doibase
		10.1038/nature13188} {\bibfield  {journal} {\bibinfo  {journal} {Nature}\
		}\textbf {\bibinfo {volume} {508}},\ \bibinfo {pages} {241} (\bibinfo {year}
		{2014})}\BibitemShut {NoStop}%
	\bibitem [{\citenamefont {Giesz}\ \emph {et~al.}(2016)\citenamefont {Giesz},
		\citenamefont {Somaschi}, \citenamefont {Hornecker}, \citenamefont {Grange},
		\citenamefont {Reznychenko}, \citenamefont {De~Santis}, \citenamefont
		{Demory}, \citenamefont {Gomez}, \citenamefont {Sagnes}, \citenamefont
		{Lemaître}, \citenamefont {Krebs}, \citenamefont {Lanzillotti-Kimura},
		\citenamefont {Lanco}, \citenamefont {Auffeves},\ and\ \citenamefont
		{Senellart}}]{giesz2016}%
	\BibitemOpen
	\bibfield  {author} {\bibinfo {author} {\bibfnamefont {V.}~\bibnamefont
			{Giesz}}, \bibinfo {author} {\bibfnamefont {N.}~\bibnamefont {Somaschi}},
		\bibinfo {author} {\bibfnamefont {G.}~\bibnamefont {Hornecker}}, \bibinfo
		{author} {\bibfnamefont {T.}~\bibnamefont {Grange}}, \bibinfo {author}
		{\bibfnamefont {B.}~\bibnamefont {Reznychenko}}, \bibinfo {author}
		{\bibfnamefont {L.}~\bibnamefont {De~Santis}}, \bibinfo {author}
		{\bibfnamefont {J.}~\bibnamefont {Demory}}, \bibinfo {author} {\bibfnamefont
			{C.}~\bibnamefont {Gomez}}, \bibinfo {author} {\bibfnamefont
			{I.}~\bibnamefont {Sagnes}}, \bibinfo {author} {\bibfnamefont
			{A.}~\bibnamefont {Lemaître}}, \bibinfo {author} {\bibfnamefont
			{O.}~\bibnamefont {Krebs}}, \bibinfo {author} {\bibfnamefont {N.~D.}\
			\bibnamefont {Lanzillotti-Kimura}}, \bibinfo {author} {\bibfnamefont
			{L.}~\bibnamefont {Lanco}}, \bibinfo {author} {\bibfnamefont
			{A.}~\bibnamefont {Auffeves}}, \ and\ \bibinfo {author} {\bibfnamefont
			{P.}~\bibnamefont {Senellart}},\ }\href {\doibase 10.1038/ncomms11986}
	{\bibfield  {journal} {\bibinfo  {journal} {Nature Communications}\ }\textbf
		{\bibinfo {volume} {7}},\ \bibinfo {pages} {11986} (\bibinfo {year}
		{2016})}\BibitemShut {NoStop}%
	\bibitem [{\citenamefont {Bennett}\ \emph {et~al.}(2016)\citenamefont
		{Bennett}, \citenamefont {Lee}, \citenamefont {Ellis}, \citenamefont
		{Farrer}, \citenamefont {Ritchie},\ and\ \citenamefont
		{Shields}}]{bennett2016}%
	\BibitemOpen
	\bibfield  {author} {\bibinfo {author} {\bibfnamefont {A.~J.}\ \bibnamefont
			{Bennett}}, \bibinfo {author} {\bibfnamefont {J.~P.}\ \bibnamefont {Lee}},
		\bibinfo {author} {\bibfnamefont {D.~J.~P.}\ \bibnamefont {Ellis}}, \bibinfo
		{author} {\bibfnamefont {I.}~\bibnamefont {Farrer}}, \bibinfo {author}
		{\bibfnamefont {D.~A.}\ \bibnamefont {Ritchie}}, \ and\ \bibinfo {author}
		{\bibfnamefont {A.~J.}\ \bibnamefont {Shields}},\ }\href {\doibase
		10.1038/nnano.2016.113} {\bibfield  {journal} {\bibinfo  {journal} {Nature
				Nanotechnology}\ }\textbf {\bibinfo {volume} {11}},\ \bibinfo {pages} {857}
		(\bibinfo {year} {2016})}\BibitemShut {NoStop}%
	\bibitem [{\citenamefont {Snijders}\ \emph {et~al.}(2016)\citenamefont
		{Snijders}, \citenamefont {Frey}, \citenamefont {Norman}, \citenamefont
		{Bakker}, \citenamefont {Langman}, \citenamefont {Gossard}, \citenamefont
		{Bowers}, \citenamefont {Exter}, \citenamefont {Bouwmeester},\ and\
		\citenamefont {Löffler}}]{snijders2016}%
	\BibitemOpen
	\bibfield  {author} {\bibinfo {author} {\bibfnamefont {H.}~\bibnamefont
			{Snijders}}, \bibinfo {author} {\bibfnamefont {J.~A.}\ \bibnamefont {Frey}},
		\bibinfo {author} {\bibfnamefont {J.}~\bibnamefont {Norman}}, \bibinfo
		{author} {\bibfnamefont {M.~P.}\ \bibnamefont {Bakker}}, \bibinfo {author}
		{\bibfnamefont {E.~C.}\ \bibnamefont {Langman}}, \bibinfo {author}
		{\bibfnamefont {A.}~\bibnamefont {Gossard}}, \bibinfo {author} {\bibfnamefont
			{J.~E.}\ \bibnamefont {Bowers}}, \bibinfo {author} {\bibfnamefont {M.~P.}\
			\bibnamefont {Exter}}, \bibinfo {author} {\bibfnamefont {D.}~\bibnamefont
			{Bouwmeester}}, \ and\ \bibinfo {author} {\bibfnamefont {W.}~\bibnamefont
			{Löffler}},\ }\href {\doibase 10.1038/ncomms12578} {\bibfield  {journal}
		{\bibinfo  {journal} {Nature Communications}\ }\textbf {\bibinfo {volume}
			{7}},\ \bibinfo {pages} {12578} (\bibinfo {year} {2016})}\BibitemShut
	{NoStop}%
	\bibitem [{\citenamefont {Astafiev}\ \emph {et~al.}(2010)\citenamefont
		{Astafiev}, \citenamefont {Zagoskin}, \citenamefont {Abdumalikov},
		\citenamefont {Pashkin}, \citenamefont {Yamamoto}, \citenamefont {Inomata},
		\citenamefont {Nakamura},\ and\ \citenamefont {Tsai}}]{astafiev2010}%
	\BibitemOpen
	\bibfield  {author} {\bibinfo {author} {\bibfnamefont {O.}~\bibnamefont
			{Astafiev}}, \bibinfo {author} {\bibfnamefont {A.~M.}\ \bibnamefont
			{Zagoskin}}, \bibinfo {author} {\bibfnamefont {A.~A.}\ \bibnamefont
			{Abdumalikov}}, \bibinfo {author} {\bibfnamefont {Y.~A.}\ \bibnamefont
			{Pashkin}}, \bibinfo {author} {\bibfnamefont {T.}~\bibnamefont {Yamamoto}},
		\bibinfo {author} {\bibfnamefont {K.}~\bibnamefont {Inomata}}, \bibinfo
		{author} {\bibfnamefont {Y.}~\bibnamefont {Nakamura}}, \ and\ \bibinfo
		{author} {\bibfnamefont {J.~S.}\ \bibnamefont {Tsai}},\ }\href {\doibase
		10.1126/science.1181918} {\bibfield  {journal} {\bibinfo  {journal}
			{Science}\ }\textbf {\bibinfo {volume} {327}},\ \bibinfo {pages} {840}
		(\bibinfo {year} {2010})}\BibitemShut {NoStop}%
	\bibitem [{\citenamefont {Hoi}\ \emph {et~al.}(2011)\citenamefont {Hoi},
		\citenamefont {Wilson}, \citenamefont {Johansson}, \citenamefont {Palomaki},
		\citenamefont {Peropadre},\ and\ \citenamefont {Delsing}}]{ho2011}%
	\BibitemOpen
	\bibfield  {author} {\bibinfo {author} {\bibfnamefont {I.-C.}\ \bibnamefont
			{Hoi}}, \bibinfo {author} {\bibfnamefont {C.~M.}\ \bibnamefont {Wilson}},
		\bibinfo {author} {\bibfnamefont {G.}~\bibnamefont {Johansson}}, \bibinfo
		{author} {\bibfnamefont {T.}~\bibnamefont {Palomaki}}, \bibinfo {author}
		{\bibfnamefont {B.}~\bibnamefont {Peropadre}}, \ and\ \bibinfo {author}
		{\bibfnamefont {P.}~\bibnamefont {Delsing}},\ }\href {\doibase
		10.1103/PhysRevLett.107.073601} {\bibfield  {journal} {\bibinfo  {journal}
			{Physical Review Letters}\ }\textbf {\bibinfo {volume} {107}},\ \bibinfo
		{pages} {073601} (\bibinfo {year} {2011})}\BibitemShut {NoStop}%
	\bibitem [{\citenamefont {Santis}\ \emph {et~al.}(2017)\citenamefont {Santis},
		\citenamefont {Antón}, \citenamefont {Reznychenko}, \citenamefont
		{Somaschi}, \citenamefont {Coppola}, \citenamefont {Senellart}, \citenamefont
		{Gómez}, \citenamefont {Lemaître}, \citenamefont {Sagnes}, \citenamefont
		{White}, \citenamefont {Lanco}, \citenamefont {Auffèves},\ and\
		\citenamefont {Senellart}}]{santis2017}%
	\BibitemOpen
	\bibfield  {author} {\bibinfo {author} {\bibfnamefont {L.~D.}\ \bibnamefont
			{Santis}}, \bibinfo {author} {\bibfnamefont {C.}~\bibnamefont {Antón}},
		\bibinfo {author} {\bibfnamefont {B.}~\bibnamefont {Reznychenko}}, \bibinfo
		{author} {\bibfnamefont {N.}~\bibnamefont {Somaschi}}, \bibinfo {author}
		{\bibfnamefont {G.}~\bibnamefont {Coppola}}, \bibinfo {author} {\bibfnamefont
			{J.}~\bibnamefont {Senellart}}, \bibinfo {author} {\bibfnamefont
			{C.}~\bibnamefont {Gómez}}, \bibinfo {author} {\bibfnamefont
			{A.}~\bibnamefont {Lemaître}}, \bibinfo {author} {\bibfnamefont
			{I.}~\bibnamefont {Sagnes}}, \bibinfo {author} {\bibfnamefont {A.~G.}\
			\bibnamefont {White}}, \bibinfo {author} {\bibfnamefont {L.}~\bibnamefont
			{Lanco}}, \bibinfo {author} {\bibfnamefont {A.}~\bibnamefont {Auffèves}}, \
		and\ \bibinfo {author} {\bibfnamefont {P.}~\bibnamefont {Senellart}},\ }\href
	{\doibase 10.1038/nnano.2017.85} {\bibfield  {journal} {\bibinfo  {journal}
			{Nature Nanotechnology}\ }\textbf {\bibinfo {volume} {12}},\ \bibinfo {pages}
		{663} (\bibinfo {year} {2017})}\BibitemShut {NoStop}%
	\bibitem [{\citenamefont {Masumoto}\ and\ \citenamefont
		{Takagahara}(2002)}]{Masumoto2002}%
	\BibitemOpen
	\bibfield  {author} {\bibinfo {author} {\bibfnamefont {Y.}~\bibnamefont
			{Masumoto}}\ and\ \bibinfo {author} {\bibfnamefont {T.}~\bibnamefont
			{Takagahara}},\ }\href
	{https://books.google.de/books/about/Semiconductor{\_}Quantum{\_}Dots.html?id=eacszlpNisgC{\&}pgis=1}
	{\emph {\bibinfo {title} {Semiconductor Quantum Dots}}}\ (\bibinfo
	{publisher} {Springer Berlin Heidelberg},\ \bibinfo {year} {2002})\ p.\
	\bibinfo {pages} {487}\BibitemShut {NoStop}%
	\bibitem [{\citenamefont {Bennett}\ \emph {et~al.}(2010)\citenamefont
		{Bennett}, \citenamefont {Patel}, \citenamefont {Skiba-Szymanska},
		\citenamefont {Nicoll}, \citenamefont {Farrer}, \citenamefont {Ritchie},\
		and\ \citenamefont {Shields}}]{bennett2010}%
	\BibitemOpen
	\bibfield  {author} {\bibinfo {author} {\bibfnamefont {A.~J.}\ \bibnamefont
			{Bennett}}, \bibinfo {author} {\bibfnamefont {R.~B.}\ \bibnamefont {Patel}},
		\bibinfo {author} {\bibfnamefont {J.}~\bibnamefont {Skiba-Szymanska}},
		\bibinfo {author} {\bibfnamefont {C.~A.}\ \bibnamefont {Nicoll}}, \bibinfo
		{author} {\bibfnamefont {I.}~\bibnamefont {Farrer}}, \bibinfo {author}
		{\bibfnamefont {D.~A.}\ \bibnamefont {Ritchie}}, \ and\ \bibinfo {author}
		{\bibfnamefont {A.~J.}\ \bibnamefont {Shields}},\ }\href {\doibase
		10.1063/1.3460912} {\bibfield  {journal} {\bibinfo  {journal} {Applied
				Physics Letters}\ }\textbf {\bibinfo {volume} {97}},\ \bibinfo {pages}
		{031104} (\bibinfo {year} {2010})},\ \bibinfo {note}
	{wOS:000280255800004}\BibitemShut {NoStop}%
	\bibitem [{\citenamefont {Főrster}(1959)}]{Forster1959}%
	\BibitemOpen
	\bibfield  {author} {\bibinfo {author} {\bibfnamefont {T.}~\bibnamefont
			{Főrster}},\ }\href {\doibase 10.1039/DF9592700007} {\bibfield  {journal}
		{\bibinfo  {journal} {Discuss. Faraday Soc.}\ }\textbf {\bibinfo {volume}
			{27}},\ \bibinfo {pages} {7} (\bibinfo {year} {1959})}\BibitemShut {NoStop}%
	\bibitem [{\citenamefont {Ficek}\ \emph {et~al.}(2002)\citenamefont {Ficek},
		\citenamefont {Tana{\'{s}}}, \citenamefont {Tanas},\ and\ \citenamefont
		{Tana{\'{s}}}}]{Ficek2002}%
	\BibitemOpen
	\bibfield  {author} {\bibinfo {author} {\bibfnamefont {Z.}~\bibnamefont
			{Ficek}}, \bibinfo {author} {\bibfnamefont {R.}~\bibnamefont {Tana{\'{s}}}},
		\bibinfo {author} {\bibfnamefont {R.}~\bibnamefont {Tanas}}, \ and\ \bibinfo
		{author} {\bibfnamefont {R.}~\bibnamefont {Tana{\'{s}}}},\ }\href {\doibase
		10.1016/S0370-1573(02)00368-X} {\bibfield  {journal} {\bibinfo  {journal}
			{Phys. Rep.}\ }\textbf {\bibinfo {volume} {372}},\ \bibinfo {pages} {369}
		(\bibinfo {year} {2002})},\ \Eprint {http://arxiv.org/abs/0302082}
	{arXiv:0302082 [quant-ph]} \BibitemShut {NoStop}%
	\bibitem [{\citenamefont {Lanco}\ and\ \citenamefont
		{Senellart}(2015)}]{Lanco2015}%
	\BibitemOpen
	\bibfield  {author} {\bibinfo {author} {\bibfnamefont {L.}~\bibnamefont
			{Lanco}}\ and\ \bibinfo {author} {\bibfnamefont {P.}~\bibnamefont
			{Senellart}},\ }in\ \href {\doibase 10.1007/978-3-319-19231-4_2} {\emph
		{\bibinfo {booktitle} {Eng. Atom-phot. Interact.}}},\ \bibinfo {editor}
	{edited by\ \bibinfo {editor} {\bibnamefont {Predojevick}}\ and\ \bibinfo
		{editor} {\bibnamefont {Mitchell}}}\ (\bibinfo  {publisher} {Springer},\
	\bibinfo {year} {2015})\ pp.\ \bibinfo {pages} {39--71},\ \Eprint
	{http://arxiv.org/abs/1502.01062} {arXiv:1502.01062} \BibitemShut {NoStop}%
	\bibitem [{\citenamefont {Armen}\ and\ \citenamefont
		{Mabuchi}(2006)}]{Armen2006}%
	\BibitemOpen
	\bibfield  {author} {\bibinfo {author} {\bibfnamefont {M.~A.}\ \bibnamefont
			{Armen}}\ and\ \bibinfo {author} {\bibfnamefont {H.}~\bibnamefont
			{Mabuchi}},\ }\href {\doibase 10.1103/PhysRevA.73.063801} {\bibfield
		{journal} {\bibinfo  {journal} {Phys. Rev. A}\ }\textbf {\bibinfo {volume}
			{73}},\ \bibinfo {pages} {063801} (\bibinfo {year} {2006})}\BibitemShut
	{NoStop}%
	\bibitem [{\citenamefont {Kimble}(1994)}]{kimble1994}%
	\BibitemOpen
	\bibfield  {author} {\bibinfo {author} {\bibfnamefont {H.~J.}\ \bibnamefont
			{Kimble}},\ }in\ \href@noop {} {\emph {\bibinfo {booktitle} {Cavity Quantum
				Electrodyn.}}},\ \bibinfo {editor} {edited by\ \bibinfo {editor}
		{\bibfnamefont {P.~R.}\ \bibnamefont {Berman}}}\ (\bibinfo  {publisher}
	{Academic Press},\ \bibinfo {year} {1994})\ p.\ \bibinfo {pages}
	{464}\BibitemShut {NoStop}%
	\bibitem [{\citenamefont {Tavis}\ and\ \citenamefont
		{Cummings}(1968)}]{Tavis1968}%
	\BibitemOpen
	\bibfield  {author} {\bibinfo {author} {\bibfnamefont {M.}~\bibnamefont
			{Tavis}}\ and\ \bibinfo {author} {\bibfnamefont {F.~W.}\ \bibnamefont
			{Cummings}},\ }\href {\doibase 10.1103/PhysRev.170.379} {\bibfield  {journal}
		{\bibinfo  {journal} {Phys. Rev.}\ }\textbf {\bibinfo {volume} {170}},\
		\bibinfo {pages} {379} (\bibinfo {year} {1968})}\BibitemShut {NoStop}%
	\bibitem [{\citenamefont {Lehmberg}(1970)}]{Lehmberg1970}%
	\BibitemOpen
	\bibfield  {author} {\bibinfo {author} {\bibfnamefont {R.~H.}\ \bibnamefont
			{Lehmberg}},\ }\href {\doibase 10.1103/PhysRevA.2.883} {\bibfield  {journal}
		{\bibinfo  {journal} {Phys. Rev. A}\ }\textbf {\bibinfo {volume} {2}},\
		\bibinfo {pages} {883} (\bibinfo {year} {1970})}\BibitemShut {NoStop}%
	\bibitem [{\citenamefont {Andrews}\ and\ \citenamefont
		{Bradshaw}(2004)}]{Andrews2004}%
	\BibitemOpen
	\bibfield  {author} {\bibinfo {author} {\bibfnamefont {D.~L.}\ \bibnamefont
			{Andrews}}\ and\ \bibinfo {author} {\bibfnamefont {D.~S.}\ \bibnamefont
			{Bradshaw}},\ }\href {\doibase 10.1088/0143-0807/25/6/017} {\bibfield
		{journal} {\bibinfo  {journal} {Eur. J. Phys.}\ }\textbf {\bibinfo {volume}
			{25}},\ \bibinfo {pages} {845} (\bibinfo {year} {2004})}\BibitemShut
	{NoStop}%
	\bibitem [{\citenamefont {Palik}(1985)}]{E1985}%
	\BibitemOpen
	\bibfield  {author} {\bibinfo {author} {\bibfnamefont {E.~D.}\ \bibnamefont
			{Palik}},\ }\href@noop {} {\emph {\bibinfo {title} {{Handbook of optical
					constants of solids}}}}\ (\bibinfo  {publisher} {Academic Press},\ \bibinfo
	{year} {1985})\BibitemShut {NoStop}%
	\bibitem [{\citenamefont {Bonifacio}\ \emph {et~al.}(1971)\citenamefont
		{Bonifacio}, \citenamefont {Schwendimann},\ and\ \citenamefont
		{Haake}}]{Bonifacio1971}%
	\BibitemOpen
	\bibfield  {author} {\bibinfo {author} {\bibfnamefont {R.}~\bibnamefont
			{Bonifacio}}, \bibinfo {author} {\bibfnamefont {P.}~\bibnamefont
			{Schwendimann}}, \ and\ \bibinfo {author} {\bibfnamefont {F.}~\bibnamefont
			{Haake}},\ }\href {\doibase 10.1103/PhysRevA.4.302} {\bibfield  {journal}
		{\bibinfo  {journal} {Phys. Rev. A}\ }\textbf {\bibinfo {volume} {4}},\
		\bibinfo {pages} {302} (\bibinfo {year} {1971})}\BibitemShut {NoStop}%
	\bibitem [{\citenamefont {Hettich}\ \emph {et~al.}(2002)\citenamefont
		{Hettich}, \citenamefont {Schmitt}, \citenamefont {Zitzmann}, \citenamefont
		{K{\"{u}}hn}, \citenamefont {Gerhardt},\ and\ \citenamefont
		{Sandoghdar}}]{Hettich2002}%
	\BibitemOpen
	\bibfield  {author} {\bibinfo {author} {\bibfnamefont {C.}~\bibnamefont
			{Hettich}}, \bibinfo {author} {\bibfnamefont {C.}~\bibnamefont {Schmitt}},
		\bibinfo {author} {\bibfnamefont {J.}~\bibnamefont {Zitzmann}}, \bibinfo
		{author} {\bibfnamefont {S.}~\bibnamefont {K{\"{u}}hn}}, \bibinfo {author}
		{\bibfnamefont {I.}~\bibnamefont {Gerhardt}}, \ and\ \bibinfo {author}
		{\bibfnamefont {V.}~\bibnamefont {Sandoghdar}},\ }\href {\doibase
		10.1126/science.1075606} {\bibfield  {journal} {\bibinfo  {journal}
			{Science}\ }\textbf {\bibinfo {volume} {298}},\ \bibinfo {pages} {385}
		(\bibinfo {year} {2002})}\BibitemShut {NoStop}%
	\bibitem [{\citenamefont {Michler}(2009)}]{Michler2009}%
	\BibitemOpen
	\bibfield  {author} {\bibinfo {author} {\bibfnamefont {P.}~\bibnamefont
			{Michler}},\ }\href {\doibase 10.1007/978-3-540-87446-1} {\emph {\bibinfo
			{title} {Single Semicond. Quantum Dots}}},\ edited by\ \bibinfo {editor}
	{\bibfnamefont {P.}~\bibnamefont {Michler}},\ NanoScience and Technology\
	(\bibinfo  {publisher} {Springer Berlin Heidelberg},\ \bibinfo {address}
	{Berlin, Heidelberg},\ \bibinfo {year} {2009})\ pp.\ \bibinfo {pages}
	{185--225},\ \Eprint {http://arxiv.org/abs/0611469} {arXiv:0611469
		[cond-mat]} \BibitemShut {NoStop}%
	\bibitem [{\citenamefont {Sipahigil}\ \emph {et~al.}(2014)\citenamefont
		{Sipahigil}, \citenamefont {Jahnke}, \citenamefont {Rogers}, \citenamefont
		{Teraji}, \citenamefont {Isoya}, \citenamefont {Zibrov}, \citenamefont
		{Jelezko},\ and\ \citenamefont {Lukin}}]{Sipahigil2014}%
	\BibitemOpen
	\bibfield  {author} {\bibinfo {author} {\bibfnamefont {A.}~\bibnamefont
			{Sipahigil}}, \bibinfo {author} {\bibfnamefont {K.~D.}\ \bibnamefont
			{Jahnke}}, \bibinfo {author} {\bibfnamefont {L.~J.}\ \bibnamefont {Rogers}},
		\bibinfo {author} {\bibfnamefont {T.}~\bibnamefont {Teraji}}, \bibinfo
		{author} {\bibfnamefont {J.}~\bibnamefont {Isoya}}, \bibinfo {author}
		{\bibfnamefont {A.~S.}\ \bibnamefont {Zibrov}}, \bibinfo {author}
		{\bibfnamefont {F.}~\bibnamefont {Jelezko}}, \ and\ \bibinfo {author}
		{\bibfnamefont {M.~D.}\ \bibnamefont {Lukin}},\ }\href {\doibase
		10.1103/PhysRevLett.113.113602} {\bibfield  {journal} {\bibinfo  {journal}
			{Phys. Rev. Lett.}\ }\textbf {\bibinfo {volume} {113}},\ \bibinfo {pages}
		{113602} (\bibinfo {year} {2014})},\ \Eprint {http://arxiv.org/abs/1406.4268}
	{arXiv:1406.4268} \BibitemShut {NoStop}%
	\bibitem [{\citenamefont {Laucht}\ \emph {et~al.}(2010)\citenamefont {Laucht},
		\citenamefont {Villas-B{\^{o}}as}, \citenamefont {Stobbe}, \citenamefont
		{Hauke}, \citenamefont {Hofbauer}, \citenamefont {B{\"{o}}hm}, \citenamefont
		{Lodahl}, \citenamefont {Amann}, \citenamefont {Kaniber},\ and\ \citenamefont
		{Finley}}]{Laucht2010}%
	\BibitemOpen
	\bibfield  {author} {\bibinfo {author} {\bibfnamefont {A.}~\bibnamefont
			{Laucht}}, \bibinfo {author} {\bibfnamefont {J.~M.}\ \bibnamefont
			{Villas-B{\^{o}}as}}, \bibinfo {author} {\bibfnamefont {S.}~\bibnamefont
			{Stobbe}}, \bibinfo {author} {\bibfnamefont {N.}~\bibnamefont {Hauke}},
		\bibinfo {author} {\bibfnamefont {F.}~\bibnamefont {Hofbauer}}, \bibinfo
		{author} {\bibfnamefont {G.}~\bibnamefont {B{\"{o}}hm}}, \bibinfo {author}
		{\bibfnamefont {P.}~\bibnamefont {Lodahl}}, \bibinfo {author} {\bibfnamefont
			{M.~C.}\ \bibnamefont {Amann}}, \bibinfo {author} {\bibfnamefont
			{M.}~\bibnamefont {Kaniber}}, \ and\ \bibinfo {author} {\bibfnamefont
			{J.~J.}\ \bibnamefont {Finley}},\ }\href {\doibase
		10.1103/PhysRevB.82.075305} {\bibfield  {journal} {\bibinfo  {journal} {Phys.
				Rev. B - Condens. Matter Mater. Phys.}\ }\textbf {\bibinfo {volume} {82}},\
		\bibinfo {pages} {075305} (\bibinfo {year} {2010})},\ \Eprint
	{http://arxiv.org/abs/0912.3685} {arXiv:0912.3685} \BibitemShut {NoStop}%
	\bibitem [{\citenamefont {Reiserer}\ and\ \citenamefont
		{Rempe}(2015)}]{Reiserer2015}%
	\BibitemOpen
	\bibfield  {author} {\bibinfo {author} {\bibfnamefont {A.}~\bibnamefont
			{Reiserer}}\ and\ \bibinfo {author} {\bibfnamefont {G.}~\bibnamefont
			{Rempe}},\ }\href {\doibase 10.1103/RevModPhys.87.1379} {\bibfield  {journal}
		{\bibinfo  {journal} {Rev. Mod. Phys.}\ }\textbf {\bibinfo {volume} {87}},\
		\bibinfo {pages} {1379} (\bibinfo {year} {2015})}\BibitemShut {NoStop}%
	\bibitem [{\citenamefont {Casabone}\ \emph {et~al.}(2015)\citenamefont
		{Casabone}, \citenamefont {Friebe}, \citenamefont {Brandst\"atter},
		\citenamefont {Sch\"uppert}, \citenamefont {Blatt},\ and\ \citenamefont
		{Northup}}]{Casabone2015}%
	\BibitemOpen
	\bibfield  {author} {\bibinfo {author} {\bibfnamefont {B.}~\bibnamefont
			{Casabone}}, \bibinfo {author} {\bibfnamefont {K.}~\bibnamefont {Friebe}},
		\bibinfo {author} {\bibfnamefont {B.}~\bibnamefont {Brandst\"atter}},
		\bibinfo {author} {\bibfnamefont {K.}~\bibnamefont {Sch\"uppert}}, \bibinfo
		{author} {\bibfnamefont {R.}~\bibnamefont {Blatt}}, \ and\ \bibinfo {author}
		{\bibfnamefont {T.~E.}\ \bibnamefont {Northup}},\ }\href {\doibase
		10.1103/PhysRevLett.114.023602} {\bibfield  {journal} {\bibinfo  {journal}
			{Phys. Rev. Lett.}\ }\textbf {\bibinfo {volume} {114}},\ \bibinfo {pages}
		{023602} (\bibinfo {year} {2015})}\BibitemShut {NoStop}%
	\bibitem [{\citenamefont {Albert}\ \emph {et~al.}(2013)\citenamefont {Albert},
		\citenamefont {Sivalertporn}, \citenamefont {Kasprzak}, \citenamefont
		{Strau{\ss}}, \citenamefont {Schneider}, \citenamefont {H{\"{o}}fling},
		\citenamefont {Kamp}, \citenamefont {Forchel}, \citenamefont {Reitzenstein},
		\citenamefont {Muljarov},\ and\ \citenamefont {Langbein}}]{Albert2013}%
	\BibitemOpen
	\bibfield  {author} {\bibinfo {author} {\bibfnamefont {F.}~\bibnamefont
			{Albert}}, \bibinfo {author} {\bibfnamefont {K.}~\bibnamefont
			{Sivalertporn}}, \bibinfo {author} {\bibfnamefont {J.}~\bibnamefont
			{Kasprzak}}, \bibinfo {author} {\bibfnamefont {M.}~\bibnamefont
			{Strau{\ss}}}, \bibinfo {author} {\bibfnamefont {C.}~\bibnamefont
			{Schneider}}, \bibinfo {author} {\bibfnamefont {S.}~\bibnamefont
			{H{\"{o}}fling}}, \bibinfo {author} {\bibfnamefont {M.}~\bibnamefont {Kamp}},
		\bibinfo {author} {\bibfnamefont {A.}~\bibnamefont {Forchel}}, \bibinfo
		{author} {\bibfnamefont {S.}~\bibnamefont {Reitzenstein}}, \bibinfo {author}
		{\bibfnamefont {E.~a.}\ \bibnamefont {Muljarov}}, \ and\ \bibinfo {author}
		{\bibfnamefont {W.}~\bibnamefont {Langbein}},\ }\href {\doibase
		10.1038/ncomms2764} {\bibfield  {journal} {\bibinfo  {journal} {Nat.
				Commun.}\ }\textbf {\bibinfo {volume} {4}},\ \bibinfo {pages} {1747}
		(\bibinfo {year} {2013})},\ \Eprint {http://arxiv.org/abs/1206.0592}
	{arXiv:1206.0592} \BibitemShut {NoStop}%
	\bibitem [{\citenamefont {Averkiev}\ \emph {et~al.}(2009)\citenamefont
		{Averkiev}, \citenamefont {Glazov},\ and\ \citenamefont
		{Poddubnyi}}]{Averkiev2009}%
	\BibitemOpen
	\bibfield  {author} {\bibinfo {author} {\bibfnamefont {N.~S.}\ \bibnamefont
			{Averkiev}}, \bibinfo {author} {\bibfnamefont {M.~M.}\ \bibnamefont
			{Glazov}}, \ and\ \bibinfo {author} {\bibfnamefont {A.~N.}\ \bibnamefont
			{Poddubnyi}},\ }\href {\doibase 10.1134/S1063776109050124} {\bibfield
		{journal} {\bibinfo  {journal} {J. Exp. Theor. Phys.}\ }\textbf {\bibinfo
			{volume} {108}},\ \bibinfo {pages} {836} (\bibinfo {year} {2009})},\ \Eprint
	{http://arxiv.org/abs/0809.3155} {arXiv:0809.3155} \BibitemShut {NoStop}%
	\bibitem [{\citenamefont {Radulaski}\ \emph {et~al.}(2017)\citenamefont
		{Radulaski}, \citenamefont {Fischer}, \citenamefont {Lagoudakis},
		\citenamefont {Zhang},\ and\ \citenamefont {Vu\ifmmode \check{c}\else
			\v{c}\fi{}kovi\ifmmode~\acute{c}\else \'{c}\fi{}}}]{Radulaski2017}%
	\BibitemOpen
	\bibfield  {author} {\bibinfo {author} {\bibfnamefont {M.}~\bibnamefont
			{Radulaski}}, \bibinfo {author} {\bibfnamefont {K.~A.}\ \bibnamefont
			{Fischer}}, \bibinfo {author} {\bibfnamefont {K.~G.}\ \bibnamefont
			{Lagoudakis}}, \bibinfo {author} {\bibfnamefont {J.~L.}\ \bibnamefont
			{Zhang}}, \ and\ \bibinfo {author} {\bibfnamefont {J.}~\bibnamefont
			{Vu\ifmmode \check{c}\else \v{c}\fi{}kovi\ifmmode~\acute{c}\else
				\'{c}\fi{}}},\ }\href {\doibase 10.1103/PhysRevA.96.011801} {\bibfield
		{journal} {\bibinfo  {journal} {Phys. Rev. A}\ }\textbf {\bibinfo {volume}
			{96}},\ \bibinfo {pages} {011801} (\bibinfo {year} {2017})}\BibitemShut
	{NoStop}%
	\bibitem [{\citenamefont {{De L{\'{e}}s{\'{e}}leuc}}\ \emph
		{et~al.}(2017)\citenamefont {{De L{\'{e}}s{\'{e}}leuc}}, \citenamefont
		{Barredo}, \citenamefont {Lienhard}, \citenamefont {Browaeys},\ and\
		\citenamefont {Lahaye}}]{DeLeseleuc2017}%
	\BibitemOpen
	\bibfield  {author} {\bibinfo {author} {\bibfnamefont {S.}~\bibnamefont {{de
					L{\'{e}}s{\'{e}}leuc}}}, \bibinfo {author} {\bibfnamefont {D.}~\bibnamefont
			{Barredo}}, \bibinfo {author} {\bibfnamefont {V.}~\bibnamefont {Lienhard}},
		\bibinfo {author} {\bibfnamefont {A.}~\bibnamefont {Browaeys}}, \ and\
		\bibinfo {author} {\bibfnamefont {T.}~\bibnamefont {Lahaye}},\ }\href
	{\doibase 10.1103/PhysRevLett.119.053202} {\bibfield  {journal} {\bibinfo
			{journal} {Phys. Rev. Lett.}\ }\textbf {\bibinfo {volume} {119}},\ \bibinfo
		{pages} {053202} (\bibinfo {year} {2017})},\ \Eprint
	{http://arxiv.org/abs/1705.03293} {arXiv:1705.03293} \BibitemShut {NoStop}%
	\bibitem [{\citenamefont {Ficek}\ \emph {et~al.}(1986)\citenamefont {Ficek},
		\citenamefont {Tanas},\ and\ \citenamefont {Kielich}}]{Ficek1986}%
	\BibitemOpen
	\bibfield  {author} {\bibinfo {author} {\bibfnamefont {Z.}~\bibnamefont
			{Ficek}}, \bibinfo {author} {\bibfnamefont {R.}~\bibnamefont {Tanas}}, \ and\
		\bibinfo {author} {\bibfnamefont {S.}~\bibnamefont {Kielich}},\ }\href
	{\doibase 10.1080/713822079} {\bibfield  {journal} {\bibinfo  {journal} {Opt.
				Acta Int. J. Opt.}\ }\textbf {\bibinfo {volume} {33}},\ \bibinfo {pages}
		{1149} (\bibinfo {year} {1986})}\BibitemShut {NoStop}%
	\bibitem [{\citenamefont {Tan}(1999)}]{Tan1999}%
	\BibitemOpen
	\bibfield  {author} {\bibinfo {author} {\bibfnamefont {S.}~\bibnamefont
			{Tan}},\ }\href {\doibase 10.1088/1464-4266/1/4/312} {\bibfield  {journal}
		{\bibinfo  {journal} {J. Opt. B Quantum Semiclass. Opt}\ }\textbf {\bibinfo
			{volume} {1}},\ \bibinfo {pages} {1} (\bibinfo {year} {1999})}\BibitemShut
	{NoStop}%
	\bibitem [{\citenamefont {DeVoe}\ and\ \citenamefont
		{Brewer}(1996)}]{DeVoe1996}%
	\BibitemOpen
	\bibfield  {author} {\bibinfo {author} {\bibfnamefont {R.~G.}\ \bibnamefont
			{DeVoe}}\ and\ \bibinfo {author} {\bibfnamefont {R.~G.}\ \bibnamefont
			{Brewer}},\ }\href {\doibase 10.1103/PhysRevLett.76.2049} {\bibfield
		{journal} {\bibinfo  {journal} {Phys. Rev. Lett.}\ }\textbf {\bibinfo
			{volume} {76}},\ \bibinfo {pages} {2049} (\bibinfo {year}
		{1996})}\BibitemShut {NoStop}%
	\bibitem [{\citenamefont {Carmichael}\ and\ \citenamefont
		{Walls}(1976)}]{Carmichael1976}%
	\BibitemOpen
	\bibfield  {author} {\bibinfo {author} {\bibfnamefont {H.~J.}\ \bibnamefont
			{Carmichael}}\ and\ \bibinfo {author} {\bibfnamefont {D.~F.}\ \bibnamefont
			{Walls}},\ }\href {\doibase 10.1088/0022-3700/9/15/524} {\bibfield  {journal}
		{\bibinfo  {journal} {J. Phys. B At. Mol. Phys.}\ }\textbf {\bibinfo {volume}
			{9}},\ \bibinfo {pages} {524} (\bibinfo {year} {1976})}\BibitemShut {NoStop}%
	\bibitem [{\citenamefont {Walls}\ and\ \citenamefont
		{Milburn}(2008)}]{Walls2008}%
	\BibitemOpen
	\bibfield  {author} {\bibinfo {author} {\bibfnamefont {D.~F.}\ \bibnamefont
			{Walls}}\ and\ \bibinfo {author} {\bibfnamefont {G.~J.}\ \bibnamefont
			{Milburn}},\ }\href {\doibase 10.1007/978-3-540-28574-8} {\emph {\bibinfo
			{title} {Quantum Opt.}}}\ (\bibinfo  {publisher} {Springer-Verlag},\ \bibinfo
	{year} {2008})\ pp.\ \bibinfo {pages} {1--425}\BibitemShut {NoStop}%
	\bibitem [{\citenamefont {Rosenblum}\ \emph {et~al.}(2011)\citenamefont
		{Rosenblum}, \citenamefont {Parkins},\ and\ \citenamefont
		{Dayan}}]{Rosenblum2011}%
	\BibitemOpen
	\bibfield  {author} {\bibinfo {author} {\bibfnamefont {S.}~\bibnamefont
			{Rosenblum}}, \bibinfo {author} {\bibfnamefont {S.}~\bibnamefont {Parkins}},
		\ and\ \bibinfo {author} {\bibfnamefont {B.}~\bibnamefont {Dayan}},\ }\href
	{\doibase 10.1103/PhysRevA.84.033854} {\bibfield  {journal} {\bibinfo
			{journal} {Phys. Rev. A}\ }\textbf {\bibinfo {volume} {84}},\ \bibinfo
		{pages} {033854} (\bibinfo {year} {2011})}\BibitemShut {NoStop}%
	\bibitem [{\citenamefont {Laussy}\ \emph {et~al.}(2008)\citenamefont {Laussy},
		\citenamefont {del Valle},\ and\ \citenamefont {Tejedor}}]{Laussy2008}%
	\BibitemOpen
	\bibfield  {author} {\bibinfo {author} {\bibfnamefont {F.~P.}\ \bibnamefont
			{Laussy}}, \bibinfo {author} {\bibfnamefont {E.}~\bibnamefont {del Valle}}, \
		and\ \bibinfo {author} {\bibfnamefont {C.}~\bibnamefont {Tejedor}},\ }\href
	{\doibase 10.1103/PhysRevLett.101.083601} {\bibfield  {journal} {\bibinfo
			{journal} {Phys. Rev. Lett.}\ }\textbf {\bibinfo {volume} {101}},\ \bibinfo
		{pages} {083601} (\bibinfo {year} {2008})}\BibitemShut {NoStop}%
	\bibitem [{\citenamefont {Carmichael}\ \emph {et~al.}(1989)\citenamefont
		{Carmichael}, \citenamefont {Brecha}, \citenamefont {Raizen}, \citenamefont
		{Kimble},\ and\ \citenamefont {Rice}}]{Carmichael1989}%
	\BibitemOpen
	\bibfield  {author} {\bibinfo {author} {\bibfnamefont {H.~J.}\ \bibnamefont
			{Carmichael}}, \bibinfo {author} {\bibfnamefont {R.~J.}\ \bibnamefont
			{Brecha}}, \bibinfo {author} {\bibfnamefont {M.~G.}\ \bibnamefont {Raizen}},
		\bibinfo {author} {\bibfnamefont {H.~J.}\ \bibnamefont {Kimble}}, \ and\
		\bibinfo {author} {\bibfnamefont {P.~R.}\ \bibnamefont {Rice}},\ }\href
	{\doibase 10.1103/PhysRevA.40.5516} {\bibfield  {journal} {\bibinfo
			{journal} {Phys. Rev. A}\ }\textbf {\bibinfo {volume} {40}},\ \bibinfo
		{pages} {5516} (\bibinfo {year} {1989})}\BibitemShut {NoStop}%
\end{thebibliography}
%

\end{document}